\documentclass[nonacm,acmsmall]{acmart}
\usepackage{xcolor}
\definecolor{bestred}{RGB}{190,0,0}
\definecolor{worstblue}{RGB}{0,70,175}
\definecolor{bestbg}{RGB}{255,226,226}
\definecolor{worstbg}{RGB}{224,235,255}
\acmJournal{TOIS}
\acmDOI{}
\usepackage{geometry}
\usepackage{booktabs}
\usepackage{graphicx}
\usepackage{amsmath}
\usepackage{tabularx}
\usepackage{multirow}
\usepackage{colortbl}
\usepackage{enumitem}
\usepackage{pifont}
\usepackage{placeins}
\newcolumntype{Y}{>{\raggedright\arraybackslash}X}
\newcolumntype{C}{>{\centering\arraybackslash}X}
\newcolumntype{L}[1]{>{\raggedright\arraybackslash}p{#1}}
\newcolumntype{M}[1]{>{\centering\arraybackslash}p{#1}}
\newcolumntype{N}[1]{>{\raggedleft\arraybackslash}p{#1}}

\newcommand{\pqbTableNote}[1]{\par\noindent\raggedright #1\par}
\newsavebox{\pqbTableWidthBox}
\title[PersonalQuery-Benchmark]{\texorpdfstring{PersonalQuery-Benchmark: Evaluating Retrieval Robustness to Personalized User Expression Differences}{PersonalQuery-Benchmark}}
\author{WENYU LIANG}
\authornote{Corresponding author.}
\affiliation{%
  \institution{Monash University}
  \city{Melbourne}
  \country{Australia}
}
\email{wlia0047@student.monash.edu}

\author{Hongzhi Yin}
\affiliation{%
  \institution{University of Queensland}
  \city{Brisbane}
  \country{Australia}
}
\email{h.yin1@uq.edu.au}

\author{TONGTONG WU}
\affiliation{%
  \institution{Monash University}
  \city{Melbourne}
  \country{Australia}
}
\email{tongtong.wu@monash.edu}

\author{Fan Liu}
\affiliation{%
  \institution{Southeast University}
  \city{Nanjing}
  \country{China}
}
\email{liufancs@gmail.com}

\author{KEXIN DAI}
\affiliation{%
  \institution{Monash University}
  \city{Melbourne}
  \country{Australia}
}
\email{hannahdai04@gmail.com}

\author{WEIQING WANG}
\affiliation{%
  \institution{Monash University}
  \city{Melbourne}
  \country{Australia}
}
\email{teresa.wang@monash.edu}
\newcommand{\pqbrepolink}{\href{https://github.com/cpxjazz123/PersonalQuery-Benchmark}%
  {\textcolor{blue}{\nolinkurl{https://github.com/cpxjazz123/PersonalQuery-Benchmark}}}}

\begin{document}
\begin{abstract}
Existing personalized information retrieval methods mainly model differences in user needs through user preferences and query intent, while paying little attention to the personalized expression differences that arise when different users express the same retrieval need. Existing real query data typically lack stable user identities, long-term expression histories, and intent-matched counterparts, making it difficult to systematically evaluate the impact of personalized expression differences on retrieval outcomes. To this end, we propose PersonalQuery-Benchmark (PQB). PQB models personalized syntactic expression and personalized writing errors from users' historical texts, constructs personalized syntactic expression queries and personalized writing-error queries, and validates benchmark quality through LLM assessment and independent human evaluation. PQB covers three Amazon product domains with 6{,}516 query instances in total, and is evaluated on five retrieval models representing lexical, learned sparse, dense bi-encoder, late-interaction, and hybrid architectures, together with two LLM-based ranking models. Results show that even when the information need is held fixed, personalized expression differences still induce measurable ranking variation; different retrieval models exhibit clearly different sensitivities to syntactic variation and writing-error variation, while higher overall retrieval effectiveness does not necessarily correspond to stronger robustness against expression variation. PQB provides a new benchmark and evaluation framework for systematically studying personalized expression robustness. The associated code and dataset are available at the following address: \pqbrepolink.
\end{abstract}
\ccsdesc[500]{Information systems~Retrieval models and ranking}
\ccsdesc[300]{Information systems~Test collections}
\keywords{personalized retrieval, syntax, robustness, benchmark}
\maketitle
\renewcommand{\shortauthors}{Wenyu Liang et al.}
\section{Introduction}
\label{sec:introduction}

Existing personalized information retrieval mainly models what different users want through user preferences, historical behavior, and query intent, and uses these signals to improve the personalized relevance of retrieval results~\cite{zhang2020dpsr,liu2023jdsearch,vemuri2023personalized,geyik2018session,joko2024personallaps,karmaker2017ecommerce,xie2020verticalsearch,yetukuri2025intentaware}. However, the differences between users are not only about the underlying information need. Even when the retrieval intent and core semantic content are kept aligned, different users can still express the same need in different linguistic forms~\cite{sundararajan2018style,stamatatos2009survey,alaofi2025demographicvariants,imasaka2024personality}. These expression differences can further affect lexical matching and query representation~\cite{benham2019boosting,penha2022robustness}, and downstream ranking~\cite{zuccon2016queryvariations}, so retrieval robustness depends not only on whether the system understands what users want, but also on whether it can stably handle how users express that need~\cite{hagen2024revisiting}.
\par\indent
We call these linguistic differences between users, which do not depend on changes in underlying intent, as personalized expression differences. In other words, personalization involves not only what users want, but also how users express the same need. In this paper, we aim to explore the impact of personalized expression difference on information retrieval methods.
Existing benchmarks, however, do not directly support this question. The personalization signals in existing personalized retrieval datasets are dominated by preference, interaction behavior, or intent differences~\cite{bassani2024personalized,vemuri2023personalized,geyik2018session}. The other real query datasets without personalization contain natural user expressions but typically lack stable user identities and sufficient historical expression evidence~\cite{sannigrahi2024synthetic,alaofi2023generative,bassani2021semantic,abe2025llmbased,zhuang2023augmenting}. Motivated by this gap, this paper focuses on the following central question: \textbf{When users express the same retrieval need differently, how robust are retrieval systems to personalized expression differences?}

\par
\begin{figure}[t]
    \centering
    \includegraphics[width=\linewidth]{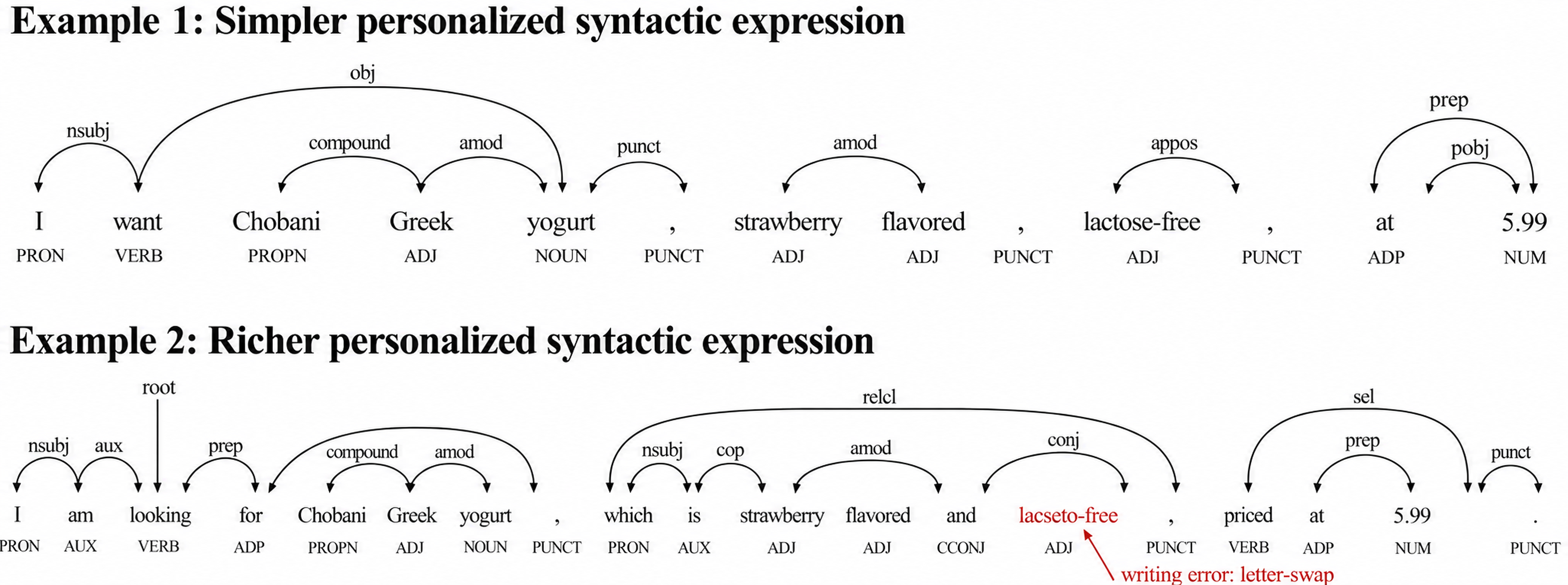}
    \caption{Example of personalized expression differences under matched product intent and core product attributes, illustrating syntactic expression and writing-error variations.}
    \Description{Two dependency parses express the same yogurt product intent. The first uses a simpler syntactic structure; the second uses a relative clause and marks the misspelling ``lacseto-free'' as a letter-swap writing error.}
    \label{fig:expression_example}
\end{figure}

\indent
To answer this question, we first need to answer which definitions can represent personalized expression. In real retrieval settings, user expression differences have already been observed as a stable phenomenon: the same user repeats similar writing habits and syntactic organizations across long-term historical texts, whereas different users differ identifiably in these respects~\cite{sundararajan2018style,stamatatos2009survey,koppel2005native,kern2013grammar}. Such differences appear both at the level of word form and in how the same semantic content is organized into a sentence. We therefore focus on two conceptually distinct and complementary dimensions of personalized expression: personalized writing errors and personalized syntactic expression. The former captures a user's long-term writing-error habits~\cite{koppel2005native,chen2017spelling,kern2013grammar}, such as letter swaps, letter repetitions, or keyboard-adjacent substitutions, and different users can show stable differences in error occurrence frequency and in which error mechanisms they prefer. The latter captures the different syntactic organizations a user can adopt when expressing the same semantic content~\cite{srikanth2003exploitingsyntax,liu2008exploringdependency,lioma2009posweighting}: even when two queries share largely the same entities, attributes, and product requirements, they can still differ in phrase organization, modifier structure, and clause construction. Fig.~\ref{fig:expression_example} illustrates both dimensions together: the personalized syntactic expression query and its corresponding personalized writing-error query convey the same core product information (Chobani Greek yogurt, strawberry flavor, lactose-free, and a price of 5.99); the former uses a more direct attribute organization, whereas the latter opens with ``I am looking for \ldots'', uses a richer clause structure, and contains the letter-swap writing error ``lacseto-free'' for ``lactose-free''.

\par\indent
Based on the two-dimensional definitions of personalized expression, this paper introduces \textbf{PersonalQuery-Benchmark (PQB)}, which evaluates how robust retrieval systems are to personalized expression differences. PQB builds on the historical user texts in Amazon Reviews and models both dimensions for each user from their historical texts. The full construction procedure is described in Sections~3.2--3.4.
We further validate the PQB along four dimensions--- \textbf{Personalized Syntactic Attribution}, \textbf{Personalized Writing-Error Consistency}, \textbf{Query Quality}, and \textbf{Evaluation Agreement}---through LLM assessment and human evaluation.

\par\indent
On this basis, PQB covers three Amazon product domains---Musical Instruments, Video Games, and Baby Products---and evaluates personalized expression robustness on five retrieval models spanning lexical, learned sparse, dense bi-encoder, late-interaction, and hybrid architectures, together with two LLM-based ranking models. The results demonstrate that, even when the underlying product need is held fixed, personalized expression differences still cause clear ranking variation; different retrieval models show different sensitivities to syntactic variation versus writing-error variation, and overall retrieval effectiveness does not map in any simple way to expression robustness.

\par\indent
The main contributions of this paper are fivefold:
\begingroup
\begin{enumerate}[topsep=0pt,itemsep=0pt,parsep=0pt,partopsep=0pt]
    \item \textbf{Personalized Expression Retrieval Setting.}
    We are the first to systematically investigate retrieval robustness to user-specific query expression differences derived from users' historical writing under a fixed information need, and we propose personalized syntactic expression and personalized writing errors as the two defining dimensions of personalized expression.
    \item \textbf{PersonalQuery-Benchmark.}
    We construct a personalized expression retrieval benchmark grounded in real users' historical expression evidence. Benchmark quality is validated along four dimensions---Personalized Syntactic Attribution, Personalized Writing-Error Consistency, Query Quality, and Evaluation Agreement---through LLM assessment and human evaluation. The benchmark covers three Amazon product domains and 6{,}516 query instances.
    \item \textbf{Personalized Expression Modeling and Query Construction.}
    The benchmark construction is grounded in users' real historical review texts. For personalized syntactic expression, PQB extracts structural rules such as dependency tuples and POS trigrams from historical expressions, learns representations that distinguish users' syntactic patterns through same-product supervised contrastive learning under a fixed product context, and fits a per-user low-rank-plus-diagonal Gaussian to capture the user's characteristic syntactic position and within-user variation. For personalized writing errors, PQB automatically extracts and aligns error annotations from historical reviews with GECToR and ERRANT, retains only errors that primarily alter surface form, and derives each user's error frequency and mechanism preference.
    \item \textbf{Retrieval Robustness Evaluation.}
    We systematically evaluate personalized expression robustness on five representative retrievers spanning lexical, learned sparse, dense bi-encoder, late-interaction, and hybrid architectures, together with two LLM-based ranking models, analyzing how syntactic variation and writing-error variation each affect ranking, and show that overall retrieval effectiveness does not correspond in any simple way to robustness to expression variation.
    \item \textbf{Personalized Expression Modeling Validation.}
    We further validate the effectiveness of personalized expression modeling through two experiments. For personalized syntactic expression, we compare the rank-one-plus-diagonal Gaussian with single-embedding representations, alternative Gaussian covariance structures, probabilistic distributions, and neighborhood-based methods. The results demonstrate our proposed rank-one-plus-diagonal Gaussian's stronger user attribution and discrimination, indicating its effectiveness in capturing user-specific syntactic patterns. For personalized writing errors, we compare user-specific error-mechanism sampling with global, uniform, character-random, and wrong-user profile baselines. Personalized sampling achieves better user attribution, demonstrating that historical writing-error profiles capture individual error-mechanism preferences and support personalized query construction in PQB.
\end{enumerate}
\endgroup

\section{Related Work}
\label{sec:related_work}

The work most closely related to PQB falls into three strands: query variation and controlled generation, the effects of writing errors on retrieval, and personalized retrieval. Prior work has shown that, even when the underlying information need stays essentially unchanged, query formulation and surface-form variation can still influence retrieval behavior; at the same time, personalized retrieval and synthetic query generation have provided the foundations for exploiting user histories and constructing controlled query variations. These strands, however, are typically studied separately---one focuses on query reformulation, another on retrieval noise, another on user preference, and another on synthetic variation---and relatively little work constructs controlled expression differences from individual users' historical language behavior while fixing the target product and the retrieval intent. PQB lies at the intersection of these strands and studies personalized expression differences as an independent evaluation variable for retrieval ranking.
\subsection{Query Variation and Controlled Query Generation}
\label{sec:related_variation}

When the underlying information need stays essentially unchanged, query formulation can still affect retrieval effectiveness. Prior work has examined verbose queries, query variability, syntactic complexity, query clarity, and query-performance prediction~\cite{bailey2016uqv100,carmel2010querydifficulty,datta2022qpp,huston2010verbose,mothe2005linguisticfeatures}, and has found that structural differences between human-written and LLM-generated queries can also change retrieval behavior~\cite{huang2026llmtexts,zendel2025linguisticretrievaldiversity}. IR research also uses POS patterns, dependency structures, and syntactic organization to describe or model query structure~\cite{lioma2009posweighting,liu2008exploringdependency,srikanth2003exploitingsyntax}; meanwhile, authorship and writing-style research shows that structural linguistic patterns can reflect individual writing characteristics~\cite{stamatatos2009survey,sundararajan2018style}. Together, these studies indicate that syntactic form is not irrelevant to retrieval behavior, yet existing retrieval studies usually treat it as query-level reformulation and rarely connect it to the long-term expression patterns of individual users.

Synthetic query generation provides a complementary route to controlled query variation. Related methods have been widely used for training-data augmentation, semantic annotation, and retrieval benchmark construction~\cite{alaofi2023generative,bassani2021semantic,filice2025datamorgana,sannigrahi2024synthetic}, and other studies have increased expression diversity through demographic attributes, personality traits, query rewriting, or term-level controls~\cite{abe2025llmbased,abuonq2025termvariants,alaofi2025demographicvariants,imasaka2024personality,mandal2019queryrewriting,zhuang2023augmenting}. These methods can generate large numbers of query variants efficiently under semantic constraints, but their variation controls usually come from prompts, generic rewriting mechanisms, or predefined attributes, and therefore need not correspond to the expression behavior that individual users develop over time. PQB likewise uses controlled generation to fix product semantics, but its personalization signals and query-selection constraints are derived from historical user texts: historical expressions are used to estimate personalized syntactic distributions and further guide candidate-query selection.
\subsection{Writing Errors and Retrieval Performance}
\label{sec:related_spelling}

Writing errors and orthographic variation can also change retrieval matching behavior. Prior work has analyzed natural spelling variation, spelling correction, and regional conventions~\cite{chari2023regionalspelling,sidiropoulos2022analysing,zhou2019spelling}, and has tested the robustness of neural retrieval models to misspellings through character-level, token-level, or adversarial perturbations~\cite{goren2019ranking,liu2023topicoriented,liu2024multigranular,pruthi2019combating,wu2023prada}. These studies show that, even when the underlying information need is preserved, local surface-form changes can still cause retrieval degradation.

Existing robustness evaluations, however, usually model writing errors as population-level or task-level perturbations, such as uniform corruption probability, fixed error mechanisms, or adversarial modifications, and rarely model long-term error frequency, mechanism preference, and context-conditioned patterns from the historical writing behavior of individual users. PQB therefore automatically identifies writing-error evidence from historical user texts, builds user-specific writing-error profiles, and applies constrained character-level editing with these profiles to the already determined personalized syntactic expression queries, thereby forming same-user personalized syntactic expression and personalized writing-error query pairs. This setting allows the effects of writing errors on retrieval outcomes to be evaluated while keeping the target product, the retrieval intent, and the original syntactic expression essentially unchanged.
\subsection{Personalized Retrieval and User Modeling}
\label{sec:related_personalized}

Personalized retrieval typically relies on user embeddings, interaction histories, session context, long-term behavior, and learning-to-rank signals to improve personalized relevance matching~\cite{ai2017hierarchical,geyik2018session,joko2024personallaps,karmaker2017ecommerce,vemuri2023personalized,yetukuri2025intentaware,zhang2020dpsr}; datasets such as JDsearch further provide real queries alongside user-interaction histories~\cite{liu2023jdsearch}. These methods use personalization signals to model the preferences, behaviors, or intents of different users, that is, they focus on what different users want and change relevance estimation or ranking accordingly.

PQB addresses a different personalization dimension. Instead of using user expression history to change the relevance target, it learns personalized expression signals from user historical texts and asks whether ranking still changes when users express the same need differently. Specifically, PQB constructs same-product cross-user personalized syntactic expression query cohorts to compare the personalized syntactic expressions of different users on the same product, and constructs same-user personalized syntactic expression and personalized writing-error query pairs to evaluate personalized writing-error perturbations. Personalization in PQB therefore serves mainly as an expression variable under study rather than as an additional user signal for improving relevance matching.

Overall, prior work has separately shown that query formulation, writing errors, user histories, and controlled query generation all affect retrieval modeling, but these directions are usually independent of one another. PQB further organizes them into a unified setting: it learns personalization signals from the historical language behavior of individual users and constructs cross-user syntactic variation and within-user writing-error variation separately, so as to systematically analyze how personalized expression differences affect retrieval outcomes.
\section{Benchmark Construction}
\label{sec:benchmark}

\begin{figure}[t]
    \centering
    \includegraphics[width=\linewidth]{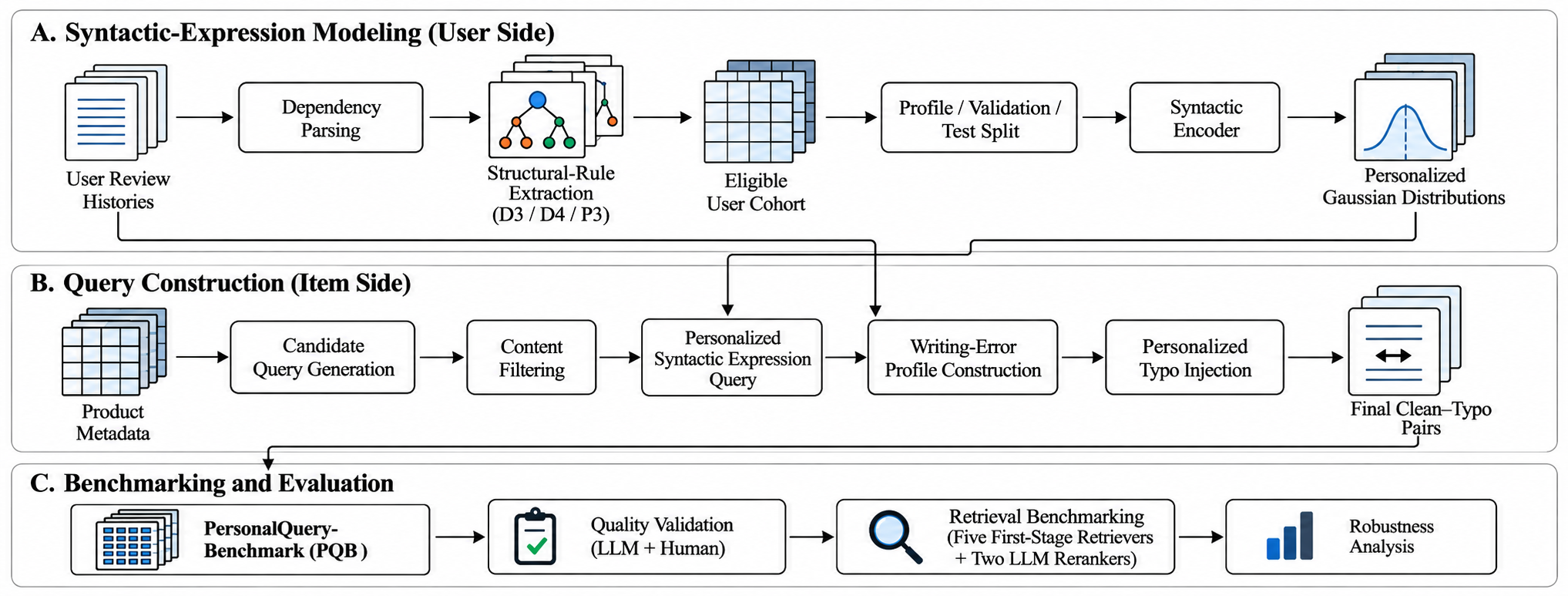}
    \caption{PQB benchmark construction pipeline from product metadata and user review histories to personalized syntactic expression queries and personalized writing-error queries.}
    \Description{A three-part flowchart shows PQB construction: user review histories are parsed into personalized syntactic distributions and writing-error profiles; product metadata drives candidate query generation and filtering; the resulting query pairs undergo quality validation, retrieval benchmarking, and analysis of retrieval outcome variation.}
    \label{fig:pqb_pipeline}
\end{figure}

PQB's construction goal is to introduce personalized variation built from real users' historical expressions while holding retrieval intent and core product semantics fixed. The construction pipeline therefore has four stages. Section~\ref{sec:dataset_construction} fixes the benchmark's data sources and sample structure and unifies product metadata with user histories as the basis for the rest of the pipeline. Section~\ref{sec:expression_quantification} learns personalized syntactic distributions from each user's historical texts. Section~\ref{sec:personalized_syntactic_query_generation} generates candidate queries under shared semantic constraints and selects one personalized syntactic expression query per user that matches the target user's syntactic expression profile. Section~\ref{sec:error_generation} constructs the corresponding personalized writing-error query based on each user's historical writing-error profile. The full benchmark construction pipeline is shown in Fig.~\ref{fig:pqb_pipeline}.

\subsection{Dataset Construction Framework}
\label{sec:dataset_construction}

As the first stage of the construction pipeline, PQB fixes the benchmark's data sources, retrieval targets, and user histories used for personalized expression modeling. PQB is built on the Amazon Reviews 2023 dataset, which provides user reviews, user--item interactions, and product metadata~\cite{hou2024bridging}. We select three product domains---Musical Instruments, Video Games, and Baby Products---shortened as Musical, Video, and Baby. PQB organizes benchmark samples by product ASIN. For each user, the products they have previously reviewed are treated as the target products the user has purchased or interacted with, and the corresponding ASINs serve as retrieval targets. PQB then links each ASIN's product metadata with the historical reviews of its associated users, so that product-side semantic information and user-side expression evidence become available together as a unified basis for personalized syntactic expression modeling and writing-error modeling.
\subsubsection{Product Constraints and User History}

After fixing the target products and their associated users, the next step is to separate the information that controls product semantics from the information that provides personalized expression evidence. The product side fixes the core product content expressed by each query. For each ASIN, PQB merges the title, description, and features into a retrieval document and selects the main product attributes from the cleaned metadata as query-generation constraints. Attribute-conditioned generation limits the semantic variation across different queries~\cite{hu2022causallens}.
Corresponding to the product-side semantic control, the user side provides personalized expression evidence. PQB aggregates the historical reviews of the users linked to each target ASIN and uses these historical texts for two modeling processes: learning each user's syntactic expression patterns and estimating each user's writing-error habits~\cite{stamatatos2009survey,vandergoot2018bleaching}.
\subsubsection{Benchmark Sample Definition}

Each benchmark row contains a target product $a$, a selected user $u$, and the personalized syntactic expression query $q_{u,a}$. For records where a valid writing-error variant is successfully generated, the corresponding personalized writing-error query $\tilde q_{u,a}$ is additionally provided. Table~\ref{tab:sample_structure} defines the fields.

\begin{table}[!ht]
\centering
\caption{Structure of a PQB benchmark sample.}
\label{tab:sample_structure}
\normalsize
\begin{tabular}{ll}
\toprule
Symbol & Meaning \\
\midrule
$a$ & Target product \\
$u$ & Target user \\
$q_{u,a}$ & Personalized syntactic expression query \\
$\tilde{q}_{u,a}$ & Personalized writing-error query \\
\bottomrule
\end{tabular}
\end{table}
\subsection{Expression-Difference Quantification}
\label{sec:expression_quantification}

\begin{figure}[t]
    \centering
    \includegraphics[width=\linewidth]{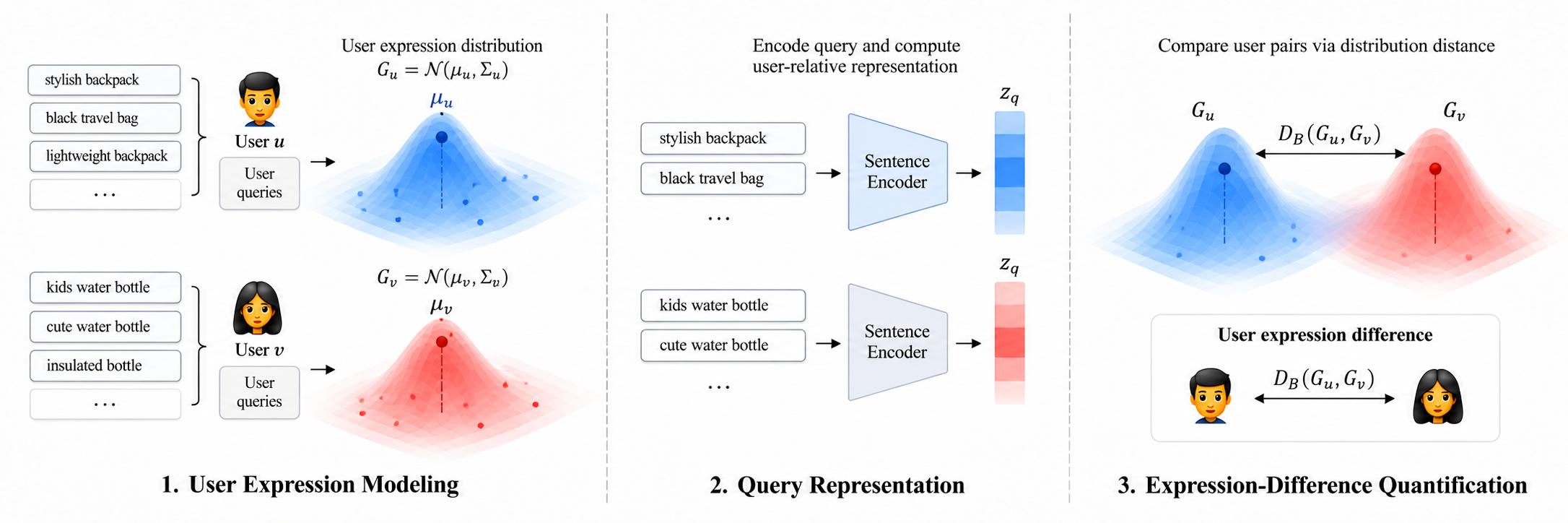}
    \caption{Personalized syntactic expression quantification: user-level expression modeling, query representation, and expression-difference measurement between user pairs.}
    \Description{A three-part flowchart shows how personalized syntactic expression differences are quantified. Part (1) models the expression distribution of one user as a Gaussian G_u = N(mu_u, Sigma_u) from that user's queries, and likewise for a second user. Part (2) encodes each candidate query with the fixed sentence encoder into a representation z_q. Part (3) compares the two fitted user Gaussians by the distribution distance D_B(G_u, G_v), which is the quantified expression difference between the two users.}
    \label{fig:syntax_quantification}
\end{figure}

After Section~3.1 fixes the target products, semantic constraints, and user histories, PQB further converts the syntactic expression patterns in each user's historical texts into comparable quantitative representations and constructs personalized syntactic distributions. The complete quantification pipeline is shown in Fig.~\ref{fig:syntax_quantification}. Concretely, Section~3.2.1 extracts structural rules from historical expressions and builds a unified syntactic feature representation; Section~3.2.2 filters eligible user--product pairs by historical sentence count; Section~3.2.3 applies same-product supervised contrastive learning, pulling the historical expressions of the same user together under a fixed product context and separating the syntactic patterns of different users, so as to learn representations that characterize personalized syntactic differences; Section~3.2.4 partitions historical expressions into profile and validation subsets; Section~3.2.5 fits a personalized low-rank-plus-diagonal Gaussian to the profile representations, capturing each user's typical syntactic position and its internal variation; Section~3.2.6 examines the validity and stability of each personalized Gaussian on held-out validation data and through repeated subsampling. Together these steps yield validated personalized syntactic distributions, which Section~3.3 then uses for personalized syntactic expression query generation.

\subsubsection{Structural Rule Extraction and Representation}

To convert the historical user texts obtained in Section~\ref{sec:dataset_construction} into a representation that mainly describes syntactic organization while minimizing the influence of specific lexical content, PQB first extracts structural rules from each historical sentence. Prior authorship and writing-style research has shown that dependency structures and POS-based patterns provide useful information for characterizing individual language expression~\cite{luyckx2008personae,nivre2016universal,sundararajan2018style}. We use three kinds of structural rules: dependency 3-tuples (D3), dependency 4-tuples with grandparent context (D4), and part-of-speech trigrams (P3). A D3 rule is written as
\[
(\text{parent POS},\text{dependency relation},\text{child POS}),
\]
for example $(\text{NOUN},\text{amod},\text{ADJ})$. A D4 rule adds the grandparent context to a D3 rule:
\[
(\text{grandparent POS},\text{parent POS},\text{dependency relation},\text{child POS}),
\]
for example $(\text{VERB},\allowbreak\text{NOUN},\allowbreak\text{amod},\allowbreak\text{ADJ})$. A P3 rule is the POS sequence of three consecutive tokens, for example $(\text{DET},\allowbreak\text{ADJ},\allowbreak\text{NOUN})$.
\par\indent
After obtaining sentence-level structural rules, PQB builds a unified global rule vocabulary so that historical expressions from different users can be mapped into the same feature space. We aggregate the unique D3, D4, and P3 rule types observed across all historical sentences and record the sentence frequency (SF) of each rule. Only rules with $SF\geq2$ are retained to reduce the sparse noise introduced by very low-frequency structural patterns.
Based on this vocabulary, each historical sentence is represented as a 21,737-dimensional sparse rule vector, whose nonzero dimensions correspond to the retained structural rules actually present in that sentence. This representation then serves as the input to the Section~3.2.3 syntax encoder.
\subsubsection{Pre-Encoder User--Product Pair Eligibility}

After Section~3.2.1 produces a unified structural representation, not every user--product pair has enough historical expressions to support reliable personalized modeling. Before training the syntax encoder, PQB therefore filters eligible user--product pairs by the historical sentence count available for each user under each product. Letting $h_{u,a}$ denote the number of historical sentences of user $u$ under product $a$, the eligibility requirement is defined as
\begin{equation}
    h_{u,a} \geq h_{\min}.
\end{equation}
This criterion depends only on the historical expression evidence observable before encoder training. It does not use learned representations, Gaussian quality, or downstream retrieval results, and therefore does not condition sample retention on later model performance.
\par\indent
To choose $h_{\min}$, we compare, across the Musical, Video, and Baby domains, the number of eligible user--product pairs at different minimum-history thresholds and the number of products that still support same-product cross-user comparison, i.e., ASINs with at least two eligible users. The results are reported in Appendix Table~\ref{tab:threshold_distribution}.
We therefore select $h_{\min}=30$ as the empirical operating point, which retains enough historical expression evidence per user--product pair while preserving as much cohort coverage as possible for same-product personalized comparison.

\par\indent
User--product pairs satisfying $h_{u,a}\geq30$ are then grouped into same-product user cohorts by parent ASIN. For each product $a$, only cohorts containing at least two eligible users are retained. This constraint directly matches the learning objective of Section~3.2.3: the syntax encoder must distinguish the syntactic expression patterns of different users under a fixed product context, where historical sentences from the same user serve as positive samples and sentences from other users under the same product provide cross-user negative samples. A product with only one eligible user cannot form such a same-product personalized comparison and is excluded from subsequent syntactic representation learning.
\subsubsection{Syntactic Representation Learning}

After Section~3.2.2 retains eligible user--product pairs with sufficient historical evidence, PQB follows the splitting strategy defined in Section~3.2.4 to partition each eligible user's historical sentences into profile and validation subsets, and uses only the profile subset for syntax encoder representation learning. After obtaining the structural-rule representations, PQB adopts same-product supervised contrastive learning: historical expressions from the same user under the same product serve as positive samples, whereas expressions from other users under that product serve as negative samples. This design encourages the representation space to reflect cross-user syntactic differences under a matched product context. For the $i$-th profile sentence of user $u$ under product $a$, the latent representation is defined as (detailed hyperparameter settings are provided in Appendix~\ref{app:syntax_encoder_config}):

\begin{equation}
z_{u,a,i}
=
f_{\theta}(c_{u,a,i})
\in
\mathbb{R}^{d}.
\label{eq:syntax_encoder}
\end{equation}

\par\indent
To preserve personalized syntactic differences within a fixed product context, training uses a supervised contrastive objective. For an anchor sentence $i$, positive samples come from other historical sentences of the same user under the same ASIN within the profile subset, while negative samples come from profile sentences of other eligible users under the same ASIN. Each eligible user is used as the anchor in turn, which lets the model learn user-specific syntactic representations under a fixed product context. All latent representations are L2-normalized, so the inner product between representations corresponds to cosine similarity.

\par\indent
Concretely, let $P(i)$ denote the positive samples belonging to the same user--product pair as anchor $i$, and let $A(i)$ denote the samples in the same ASIN cohort, other than the anchor, that participate in the contrastive comparison. Training uses

\begin{equation}
\mathcal{L}
=
\sum_{i=1}^{N}
-
\frac{1}{|P(i)|}
\sum_{p\in P(i)}
\log
\frac{
\exp(z_i^\top z_p/\tau)
}{
\sum_{a\in A(i)}
\exp(z_i^\top z_a/\tau)
}.
\label{eq:supcon}
\end{equation}

\par\indent
Through this cohort-based sampling, the encoder pulls historical expressions from the same user's profile subset closer together within a product and pushes away syntactic representations from other users. Section~3.2.4 describes how the profile and validation subsets are formed and how they are used in Gaussian fitting and validation.
\subsubsection{Profile--Validation Split}
\label{sec:profile_validation_split}

For each eligible user--product pair, PQB uses a deterministic split based on the sentence identifiers to divide historical sentences 80\% / 20\% into profile and validation subsets. The profile subset is used for syntax encoder learning and personalized Gaussian fitting; the validation subset does not participate in parameter estimation and is reserved for the held-out validity checks of Section~3.2.6. This fixed split guarantees that the same sentence never enters both fitting and validation and lets all experiments reproduce the same held-out samples. Because the eligibility threshold is $h_{u,a}\geq30$, every pair's validation subset contains at least five held-out representations.
\subsubsection{Per-User Low-Rank-plus-Diagonal Gaussian Modeling}
\label{sec:gaussian_modeling}

After Section~3.2.4 yields the profile representations, PQB aggregates the profile representations of each user across all of that user's eligible product contexts and fits a personalized syntactic distribution at the user level.
\par\indent
A single embedding vector can represent a user's typical syntactic style but cannot describe the variation among the different syntactic expressions that the same user may employ. Prior studies have adopted Gaussian embeddings to represent variability beyond fixed vector representations~\cite{vilnis2015gaussian,jiang2019convolutional}, while distribution-based personalized recommendation has used Gaussian means and covariances to characterize user preferences and relationships between latent dimensions~\cite{dong2023dual}. Motivated by these approaches, PQB adopts a Gaussian distribution as a tractable approximation of each user's historical syntactic representations, rather than assuming that these representations strictly follow a Gaussian distribution. The mean characterizes the user's typical syntactic style, while the covariance captures within-user variation. This representation also provides a unified basis for Gaussian likelihood scoring and distance-based query selection in Section~3.3. The structure of the covariance determines how much of the within-user variation this distribution can retain: a full covariance matrix is the most general choice, but each user provides only a limited number of historical expressions, so estimating it directly introduces substantial estimation noise, whereas keeping only the dimension-wise variances is more stable yet discards the correlated variation across latent dimensions. The rank-one-plus-diagonal structure splits the difference: a low-rank component along a principal direction captures the correlated variation among dimensions, and diagonal residuals model the remaining dimension-wise variation, so the within-user variation structure is retained at a tractable estimation complexity.

\par\indent
For user $u$, let the profile subset contain $n_u$ latent syntactic representations:

\begin{equation}
Z_u^{\mathrm{profile}}
=
\{z_{u,1},\ldots,z_{u,n_u}\},
\qquad
z_{u,i}\in\mathbb{R}^{16}.
\label{eq:profile_set}
\end{equation}

\par\indent
Compared with a single style vector, a distributional representation captures not only the typical location of a user's historical expressions but also their within-user variation. PQB therefore uses the empirical mean to describe the central position of the user's historical syntactic expressions in the latent space:

\begin{equation}
\mu_u
=
\frac{1}{n_u}
\sum_{i=1}^{n_u} z_{u,i}.
\label{eq:user_mean}
\end{equation}

\par\indent
Motivated by low-rank probabilistic covariance modeling~\cite{tipping1999probabilistic}, PQB adopts the rank-one-plus-diagonal covariance structure:

\begin{equation}
\Sigma_u
=
\lambda_u v_u v_u^\top
+
\operatorname{diag}(\sigma_u^2),
\label{eq:lowrank_diag_cov}
\end{equation}

where $v_u$ denotes the main variation direction of the user's syntactic representations, $\lambda_u$ is the variance contribution along that direction, and $\sigma_u^2$ is the residual variance along each latent dimension. These parameters are estimated from the empirical covariance $C_u$ of each user's profile representations. Specifically, $v_u$ is the eigenvector associated with the largest eigenvalue $\lambda_u$ of $C_u$, and the dimension-wise residual variance is obtained from the diagonal of the covariance remaining after subtracting the rank-1 component,
\begin{equation}
\sigma_u^2
=
\operatorname{diag}
\left(
C_u-\lambda_u v_u v_u^\top
\right),
\end{equation}
with each residual variance lower-bounded by $10^{-4}$ for numerical stability.

\par\indent
Based on these parameters, the personalized syntactic distribution of user $u$ is represented as

\begin{equation}
z \mid u
\mathrel{\approx}
\mathcal{N}(\mu_u,\Sigma_u).
\label{eq:user_gaussian}
\end{equation}

Here, $\mu_u$ captures the user's characteristic syntactic position, while $\Sigma_u$ describes the principal variation direction and dimension-wise residual variation. The fitted distribution is subsequently evaluated for numerical validity, held-out calibration, and estimation stability in Section~3.2.6.
\subsubsection{Gaussian Validity Filter}
\label{sec:gaussian_validity}

After Section~3.2.5 produces per-user personalized Gaussians, PQB further applies held-out validation data and runs three sequential checks on each Gaussian profile: (1)~\textbf{numerical validity}, (2)~\textbf{held-out calibration}, and (3)~\textbf{repeated-sampling stability}. Only profiles that pass all three checks are retained and used for the personalized syntactic expression query generation in Section~3.3.
\begingroup
\begin{enumerate}[topsep=2pt,itemsep=2pt,partopsep=0pt]
\item \textbf{Numerical validity.} Checks whether the Gaussian parameters and the validation Mahalanobis distances can be stably computed: the rank-1 variance and all diagonal residual variances must be positive, the Gaussian log-determinant must be finite, and the squared Mahalanobis distances for the validation representations must be computable. This step primarily removes covariance degenerations and other numerical failures.
\item \textbf{Held-out calibration.} To evaluate how well a user's Gaussian fits unseen syntactic expressions, we compute the Mahalanobis distances of held-out validation representations and require that at least 80\% of each user's validation samples satisfy $D^2\leq\chi^2_{16,0.95}=26.296$.
\item \textbf{Repeated-sampling stability.} To evaluate the stability of user Gaussian distributions against variations in historical samples, PQB repeatedly performs random subsampling and refits the Gaussians to examine their numerical validity, held-out distance ranking, and acceptance-region consistency. Specific filtering thresholds and stage-wise filtering results are provided in Appendix Table~\ref{tab:gaussian_filter_config} and Table~\ref{tab:gaussian_filter}. Empirical results show that the vast majority of Gaussian profiles satisfy the numerical validity requirement, with held-out calibration serving as the primary filtering stage, while repeated-sampling stability further removes distributions that are sensitive to historical sample variations. All three domains exhibit a consistent filtering trend, indicating that this filtering procedure improves the stability of the final Gaussian profiles while retaining high user coverage.
\end{enumerate}
\endgroup
\subsection{Personalized Syntactic Expression Query Generation}
\label{sec:personalized_syntactic_query_generation}

After Section~3.2 yields fitted syntactic distributions, PQB constructs history-informed queries under matched product attributes. Section~3.3.1 describes the frozen candidate pool and its shared product constraints. Section~3.3.2 encodes candidates with the fixed syntax encoder and applies exact Gaussian scoring and feasibility-aware pair-conditional selection.

\subsubsection{Content-Controlled Candidate Query Generation}

Section~3.3.1 first constructs a candidate query pool under strict lexical and semantic constraints. For each target product, PQB cleans the product metadata by removing missing, duplicate, and unsuitable attributes, and normalizes attribute values with format anomalies. It then fixes five core retrieval terms for each product, which serve as shared lexical constraints for all candidate queries of that product. Each query must fully retain these five core retrieval terms, with each term appearing exactly once; deletion, repetition, replacement, or synonymous rewriting is not allowed. Queries may not introduce additional product attributes or new product semantic content. Variation between queries is allowed only in function words, connective expressions, modification structures, and clause organization, i.e., in syntactic realization.
\par\indent
PQB uses Qwen2.5-0.5B-Instruct~\cite{yang2024qwen} to generate first-person shopper queries under a unified generation configuration across the three product domains. The above lexical and semantic constraints guide generation and are enforced through post-generation content validation. PQB retains only queries in which all five core retrieval terms appear exactly once and no additional product semantic content is introduced. For products with insufficient valid candidates, PQB continues iterative candidate generation and adds validated queries to the candidate pool. Subsequently, the syntax encoder from Section~3.2 encodes the candidate queries, and personalized syntactic expression query selection is completed in Section~3.3.2. Detailed generation parameters are provided in Appendix Table~\ref{tab:candidate_generation}.
\par\indent
To keep the comparison within one product about expression form rather than meaning, PQB further constrains the semantic similarity of the retained query pairs. For two queries $q_1$ and $q_2$ of the same target product, $\operatorname{SemanticSim}(q_1,q_2)$ denotes the cosine similarity between their sentence embeddings computed with the all-mpnet-base-v2 encoder~\cite{reimers2019sentence}, and PQB retains only query pairs that satisfy $\operatorname{SemanticSim}(q_1,q_2)\geq 0.9$. The constraint is enforced during construction, so the query pairs evaluated in Section~4 differ in syntactic realization rather than in semantic content, with the five core retrieval terms remaining the shared lexical basis of both queries.
\subsubsection{Gaussian-Based Personalized Syntactic Expression Query Selection}
\label{sec:gaussian_query_selection}
PQB uses the personalized syntactic distributions learned in Section~3.2 to select representative queries for different users from the candidate query pool of the same product. Unlike random sampling from the user population, PQB focuses on user pairs with clear syntactic expression differences, in order to construct challenging query samples that can evaluate retrieval models' sensitivity to cross-user expression variation. To this end, the query selection process jointly considers the agreement between candidate queries and the users' syntactic distributions, as well as the differences between different users' syntactic distributions.

For a product $a$, let $\mathcal{U}_a$ be the set of users with valid Gaussian profiles and let $\mathcal{Q}_a$ be the candidate query set constructed in Section~3.3.1. All candidate queries are mapped by the trained syntax encoder into a 16-dimensional representation space, giving the syntactic representations $z_q$. For a user $u$, the personalized syntactic distribution is represented as $G_u=\mathcal{N}(\mu_u,\Sigma_u)$. The log-probability density of a candidate query $q$ under this distribution is defined as
\begin{equation}
\log p_u(z_q) =
-\frac{1}{2}\left[
(z_q-\mu_u)^\top\Sigma_u^{-1}(z_q-\mu_u)
+\log|\Sigma_u|+16\log(2\pi)
\right].
\end{equation}

To select queries from the candidate pool of the same product that match the syntactic expression characteristics of different users, PQB filters candidates by their agreement with the Gaussian distributions of two users. For a user pair $\{u,v\}$, the candidate queries of user $u$ must lie inside the acceptance region of its own Gaussian, outside the acceptance region of user $v$, and satisfy a cross-user log-density margin. Their eligible query set is defined as
\begin{equation}
\begin{aligned}
\mathcal{A}_u(v)=\{q\in\mathcal{Q}_a:\;&D_u^2(z_q)\leq\tau,\\
&D_v^2(z_q)>\tau,\\
&\log p_u(z_q)-\log p_v(z_q)\geq 1\},
\end{aligned}
\end{equation}
where
\begin{equation}
D_u^2(z_q)=
(z_q-\mu_u)^\top\Sigma_u^{-1}(z_q-\mu_u)
\end{equation}
denotes the squared Mahalanobis distance of a candidate query under user $u$'s Gaussian distribution, and $\tau=\chi^2_{16,0.95}=26.296$ is the Gaussian reference threshold. The same filtering rule applies symmetrically to user $v$, yielding its eligible query set $\mathcal{A}_v(u)$.

For each user, PQB selects the query with the highest Gaussian log-probability density from its eligible query set:
\begin{equation}
\hat q_{u,a}^{(u,v)}=
\underset{q\in \mathcal{A}_u(v)}{\arg\max}
\Bigl(\log p_u(z_q),\ \log p_u(z_q)-\log p_v(z_q),\ -i(q)\Bigr),
\end{equation}
where $i(q)$ denotes the index of the candidate query in the original candidate pool, and the triple is maximized lexicographically. The rule first prefers the query that best matches the target user's syntactic distribution, then uses the cross-user density margin and the candidate index to resolve ties.

PQB then enumerates the user pairs under each product. A pair is feasible when both users have at least one eligible query and the finally selected queries have different texts and different source indices. The feasible user pair set for product $a$ is defined as
\begin{equation}
\begin{aligned}
\mathcal{F}_a=\bigl\{\{u,v\}\subseteq\mathcal{U}_a:\;&\mathcal{A}_u(v)\neq\emptyset,\\
&\mathcal{A}_v(u)\neq\emptyset,\\
&\hat q_{u,a}^{(u,v)}\neq\hat q_{v,a}^{(u,v)}
\bigr\},
\end{aligned}
\end{equation}
where the final condition states that the selected query texts differ, and their source indices are also guaranteed to differ during selection.

To construct evaluation samples with large inter-user syntactic expression differences, PQB selects the user pair with the largest distance between Gaussian distributions among the feasible pairs:
\begin{equation}
\{u_a^*,v_a^*\}=
\underset{\{u,v\}\in\mathcal{F}_a}{\arg\max}
D_{\mathrm B}(G_u,G_v),
\end{equation}
where $D_{\mathrm B}$ denotes the Bhattacharyya distance:
\begin{equation}
\begin{aligned}
D_{\mathrm B}(G_u,G_v)={}&\frac18(\mu_u-\mu_v)^\top
\bar\Sigma^{-1}(\mu_u-\mu_v)\\&+\frac12\log
\frac{|\bar\Sigma|}{\sqrt{|\Sigma_u||\Sigma_v|}},
\end{aligned}
\end{equation}
and $\bar\Sigma=(\Sigma_u+\Sigma_v)/2$. When several feasible user pairs share the same distributional distance, selection is deterministic according to the sorted user identifiers.

The Bhattacharyya distance simultaneously considers the centers and covariance differences of the users' syntactic distributions, and can therefore measure the distinguishability between different users' expression distributions. By maximizing this distance, PQB preferentially retains feasible user pairs with larger distributional differences, enhancing the distinguishability of evaluation samples at the syntactic expression level.

Through the above procedure, PQB first screens user pairs satisfying the personalized query matching constraints, then selects the two users with the largest Gaussian distributional distance, and respectively selects from the candidate query pool the query that best matches each user's personal syntactic distribution. These queries reflect different users' syntactic expression features under shared product retrieval content, providing a unified product context for subsequent analysis of cross-user retrieval variability.

\begin{figure}[t]
    \centering
    \includegraphics[width=\linewidth]{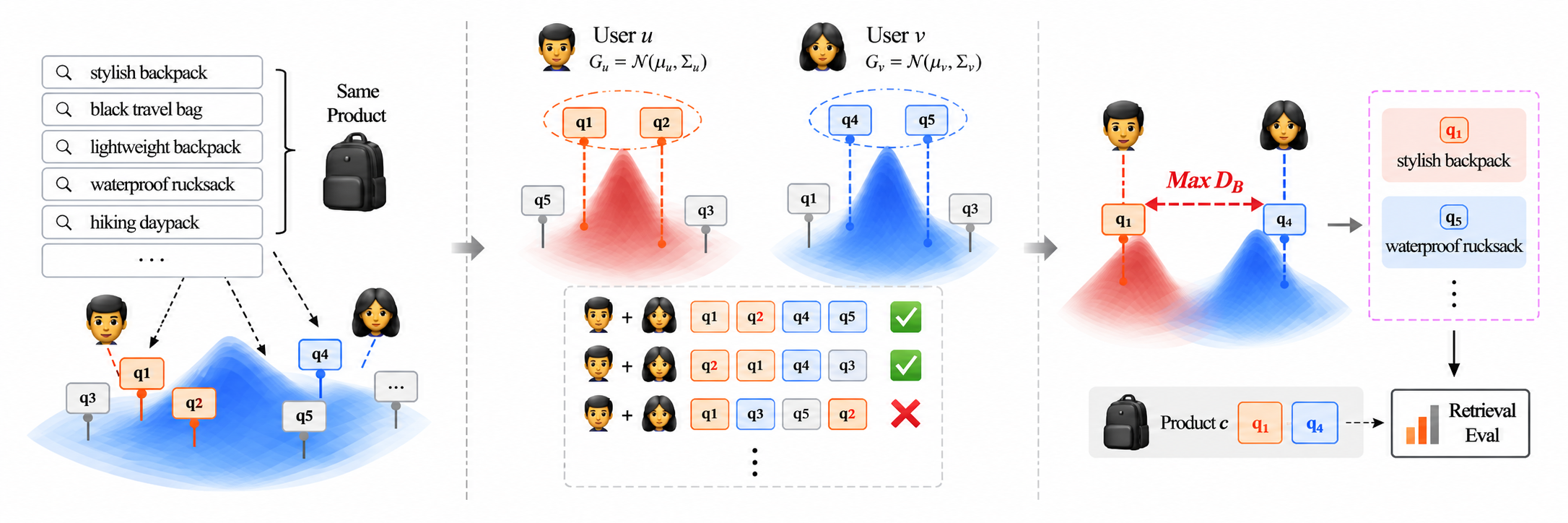}
    \caption{Gaussian-based personalized syntactic expression query selection from a shared candidate pool: pair-conditioned eligibility filtering, feasible user-pair enumeration, and Bhattacharyya-maximizing pair selection.}
    \Description{A four-part flowchart shows personalized syntactic expression query selection for one product. A candidate query pool with 16-dimensional syntactic representations is filtered per user pair against two personalized Gaussians, marking compatible and incompatible queries and stating the eligible query set conditions. Feasible user pairs are enumerated and ranked by Bhattacharyya distance; the maximum-distance pair is selected, and one representative query per selected user is assigned to the same product for retrieval evaluation.}
    \label{fig:gaussian_query_selection}
\end{figure}

As shown in Fig.~\ref{fig:gaussian_query_selection}, PQB first filters candidate queries according to the users' Gaussian syntactic distributions, then maximizes the Bhattacharyya distance among the user pairs satisfying the query feasibility constraints, and finally selects a representative personalized syntactic expression query for each of the two users of the same product.

\subsection{Personalized Writing-Error Generation}
\label{sec:error_generation}

After Section~3.3 constructs the personalized syntactic expression queries, PQB further uses the writing-error characteristics in the users' historical reviews to generate personalized character-level perturbations for the queries. The process first extracts writing errors and their syntactic contexts from the users' historical reviews to build personalized error profiles, then applies local character modifications to these queries according to the users' historical transformation records and error-mechanism preferences, forming personalized writing-error queries for evaluating changes in retrieval performance.

\subsubsection{Personalized Writing-Error Profile Construction}
\label{sec:error_profile}

PQB extracts writing-error characteristics from the users' historical reviews to construct personalized error profiles. Specifically, GECToR~\citep{omelianchuk2020gector} first produces the corrected forms of the texts, and ERRANT~\citep{bryant2017automatic} then aligns the original and corrected texts to identify and annotate the edit operations. For the extracted historical edits, PQB records the error type, the transformation form and the associated syntactic context, and counts the historical transformations that are available for character-level perturbation.
\par\indent
Unlike continuous syntactic representations, writing errors consist of discrete character-level editing operations. Their personalized characteristics involve not only how frequently users make errors, but also which error mechanisms they prefer and how these mechanisms relate to syntactic contexts. A shared global error distribution overlooks individual differences in mechanism preferences, whereas a context-independent user distribution cannot capture the relationship between writing errors and local syntactic structures. Meanwhile, fine-grained contexts may contain only limited historical error evidence for individual users. PQB therefore adopts a user-specific, context-conditioned error-mechanism distribution estimated from historical frequencies, while retaining each user's overall mechanism preferences and observed character transformations.
To characterize a user's writing-error preferences under different syntactic structures, PQB further counts the relation between error mechanisms and syntactic contexts. For user $u$, let $n_u(e,r)$ denote the number of historical occurrences of error mechanism $e$ in syntactic context $r$; the context-conditioned error-mechanism distribution is then defined as
\begin{equation}
P_u(e\mid r)
=
\frac{n_u(e,r)}
{\sum_{e' \in \mathcal{E}} n_u(e', r)}.
\end{equation}
where $\mathcal{E}$ denotes the set of error mechanisms. This distribution describes the relative frequency of different error mechanisms for the user in a given syntactic context. In parallel, PQB aggregates the user's historical edits to compute the overall character-level error-mechanism preferences and historical transformation frequencies. A user's writing-error profile therefore consists of the historical edits, the global error-mechanism preferences and the context-conditioned error distribution, and is used for the personalized writing-error query generation in Section~\ref{sec:error_query_generation}. The per-domain user profile statistics are reported in Appendix Table~\ref{tab:writing-error-profile}.
\subsubsection{Personalized Writing-Error Query Generation}
\label{sec:error_query_generation}

After the personalized writing-error profiles are obtained, PQB perturbs the personalized syntactic expression queries selected in Section~3.3 at the character level. For user $u$ and the corresponding product $a$, let the personalized syntactic expression query be $q_{u,a}$ and the generated personalized writing-error query be $\tilde{q}_{u,a}$. The process uses the user's historical error mechanisms and their context information to modify local characters of the query, constructing paired queries for evaluating the impact of writing errors on retrieval outcomes. For each personalized syntactic expression query, PQB ranks the alphabetic tokens of at least three characters by the user's historical character-level error rate and determines the candidate edit positions. At each selected position, PQB first attempts to reuse a character transformation observed in the user's historical reviews. Transformation matching proceeds from the most specific context to the least specific, backing off from the full syntactic context through dependency and POS information, coarse-grained context, and token-level context. When no applicable historical transformation exists, PQB samples an error type from the frequency distribution of the user's overall character-level error mechanisms and generates the corresponding character modification. To control the perturbation scope, PQB attempts to modify at most three distinct token positions per query and constrains the edit distance and the query length change. To monitor the semantic effect of these edits, PQB constrains the semantic similarity between each personalized syntactic expression query and its writing-error counterpart: a generated writing-error variant is rejected when its semantic similarity to the original query falls below $0.9$, where $\operatorname{SemanticSim}(q_{u,a},\tilde{q}_{u,a})$ is computed with the same all-mpnet-base-v2 encoder~\cite{reimers2019sentence}. The detailed per-domain construction statistics are reported in Appendix Table~\ref{tab:writing_error_variant_construction}.
\section{Retrieval Benchmark Evaluation}
\label{sec:evaluation}

This section evaluates, through five research questions, how personalized syntactic expression differences and writing errors affect retrieval results, and validates the constructed components of PQB. RQ1 first examines whether personalized syntactic expressions from different users correspond to different retrieval outcomes under a fixed retrieval target. RQ2 evaluates the effect of personalized writing errors on retrieval hits and ranking performance, and RQ3 validates benchmark quality in terms of personalized characteristics, query quality, and evaluation agreement. RQ4 compares the adopted user-specific syntactic representation with alternative embedding- and distribution-based models on held-out user discrimination and expression coverage, and RQ5 examines whether the constructed writing-error profiles identify users better than non-personalized sampling baselines. By applying a unified paired evaluation protocol across different retrieval systems, this section examines whether the above retrieval-result changes are consistent across systems.
\subsection{Experimental Setup}
\label{sec:experimental_setup}

This study revolves around the following five research questions:

\begingroup
\begin{enumerate}[topsep=2pt,itemsep=2pt,partopsep=0pt]
\item \textbf{RQ1. Personalized Syntactic Expression Differences and Retrieval Robustness.} Under a fixed retrieval target, do different personalized syntactic expressions produce retrieval-result differences? Is this phenomenon stable across different retrieval systems?
\item \textbf{RQ2. Personalized Writing Errors and Retrieval Robustness.} How do personalized writing errors affect retrieval hits and ranking performance? Is this effect stable across different retrieval systems?
\item \textbf{RQ3. Benchmark Quality Validation.} Do the queries generated by PQB exhibit good linguistic quality and identifiable personalized syntactic characteristics? Are the generated writing errors consistent with users' historical error behavior, and is the agreement between automatic and human evaluation reliable?
\item \textbf{RQ4. User-Specific Syntactic Modeling Comparison.} Compared with other user-specific syntactic representations, does the adopted rank-one-plus-diagonal Gaussian model users' syntactic expressions more effectively and achieve better results in user attribution and user discrimination?
\item \textbf{RQ5. Personalized Writing-Error Profile Evaluation.} Compared with global, uniform, character-random, and wrong-user sampling, does the proposed personalized writing-error mechanism sampling capture users' specific writing-error patterns more effectively and achieve better user attribution performance?
\end{enumerate}
\endgroup
\subsubsection{Retrieval Systems and Evaluation Protocol}
\label{sec:retrieval_systems_protocol}

To evaluate the retrieval-result changes corresponding to different user query expressions, we adopt five first-stage retrieval models with different representation and matching mechanisms: BM25~\cite{robertson2009bm25}, SPLADE~\cite{formal2021splade}, GTE-base~\cite{li2023gte}, ColBERTv2~\cite{santhanam2022colbertv2}, and BGE-M3 Hybrid~\cite{chen2024bgem3}, together with two LLM-based reranking models: Llama-3.1-8B~\cite{meta2024llama3} and Qwen3-Reranker-8B~\cite{yang2025qwen3}. These models cover lexical retrieval, learned sparse retrieval, dense bi-encoder, late interaction, hybrid retrieval, and LLM-based reranking. In particular, the two LLM-based ranking models allow us to examine whether personalized syntactic expression variations and writing-error variations affect retrieval outcomes consistently across general-purpose and specialized LLM ranking models. Detailed model and parameter configurations are provided in Appendix Table~\ref{tab:model_config}.

\subsubsection{Evaluation Metrics}
\label{sec:evaluation_metrics}

We measure retrieval-result differences corresponding to query changes from two aspects: the hit status and the reciprocal rank of the target product. Both aspects are computed for each retrieval model, and a model whose outcomes vary less under personalized expression differences is the more robust one; the reported differences are therefore read as the robustness behavior of individual models. For retrieval system $m$, let $r_m(q)$ be the retrieval rank of the target product for query $q$, with $r_m(q)=\infty$ when the target product is not retrieved. We first define the top-$K$ hit indicator:
\begin{equation}
H_{m,K}(q)
=
\mathbb{I}[1\leq r_m(q)\leq K].
\end{equation}
Here, $H_{m,K}(q)=1$ indicates that the target product appears in the top-$K$ retrieval results, and 0 otherwise.

\begingroup
\begin{enumerate}[topsep=2pt,itemsep=2pt,partopsep=0pt]
\item \textbf{Flip@$K$.} This metric measures whether two different queries produce different top-$K$ hit statuses under the same retrieval target, considering only query pairs in which at least one query retrieves the target product. For the $N$ query pairs in the evaluation set, it is defined as:
\begin{equation}
\mathrm{Flip}_m@K
=
\frac{
\sum_{i=1}^{N}
\mathbb{I}
\left[
H_{m,K}(q_{i,1})
\neq
H_{m,K}(q_{i,2})
\right]
}{
\sum_{i=1}^{N}
\mathbb{I}
\left[
H_{m,K}(q_{i,1})
\lor
H_{m,K}(q_{i,2})
\right]
}.
\end{equation}
Here, $q_{i,1}$ and $q_{i,2}$ denote the two queries corresponding to the same target product. Query pairs in which neither query retrieves the target product within the top-$K$ are excluded from the denominator. A higher Flip@$K$ indicates a higher proportion of different target-product hit statuses among query pairs with at least one successful retrieval. When the denominator is zero, Flip@$K$ is reported as N/A. We use $K\in\{1,5,10,20\}$ to evaluate retrieval-result differences at different retrieval depths.
\item \textbf{Mean $RR_{\mathrm{std}}$.} To measure the difference in target-product rank between two queries, we first define the reciprocal rank within the top 200 retrieval results:
\begin{equation}
RR_{m,200}(q)=
\begin{cases}
1/r_m(q), & 1\leq r_m(q)\leq 200,\\
0, & \text{otherwise}.
\end{cases}
\end{equation}
We then compute the sample standard deviation $\operatorname{Std}_{sample}(RR_1,RR_2)=\frac{|RR_1-R_2|}{\sqrt{2}}$ of the reciprocal ranks of the two queries for each query pair and average it over all query pairs:
\begin{equation}
\mathrm{Mean\ RR}_{\mathrm{std},m}
=
\frac{1}{N}
\sum_{i=1}^{N}
\operatorname{Std}_{sample}
\left(
RR_{m,200}(q_{i,1}),
RR_{m,200}(q_{i,2})
\right).
\end{equation}
A higher value indicates a larger reciprocal-rank difference for the same target product under the two queries. Unlike Flip@$K$, which focuses on hit status at a given retrieval depth, Mean $RR_{\mathrm{std}}$ further measures the rank change of the target product.
\item \textbf{Deg@$K$.} To measure the effect of writing errors on retrieval performance, let $q_i$ be the personalized syntactic expression query and $\tilde q_i$ its corresponding personalized writing-error query. Deg@$K$ is defined as the net hit-rate degradation between the two query conditions:
\begin{equation}
\mathrm{Deg}_m@K
=
\frac{1}{N}
\sum_{i=1}^{N}
\left[
H_{m,K}(q_i)
-
H_{m,K}(\tilde q_i)
\right].
\end{equation}
When Deg@$K$ is positive, writing errors reduce the overall top-$K$ hit rate of the target product; when it is negative, the overall hit rate improves. Because this metric simultaneously contains hit-to-miss and miss-to-hit changes, it reflects the net retrieval-performance change rather than a one-directional hit loss.
\item \textbf{MRR Degradation.} In addition to hit-rate change, we further use the mean reciprocal rank to measure the rank degradation corresponding to writing errors. Let $RR_m(q)$ be the reciprocal rank of the target product within the effective ranking range of the corresponding retrieval configuration, taking zero on a miss. The MRR change between the personalized syntactic expression query and the personalized writing-error query is defined as:
\begin{equation}
\Delta MRR_m
=
\frac{1}{N}
\sum_{i=1}^{N}
\left[
RR_m(\tilde q_i)
-
RR_m(q_i)
\right].
\end{equation}
To make a positive value correspond to retrieval-performance degradation, we further define:
\begin{equation}
\mathrm{MRR}_{\mathrm{deg},m}
=
-\Delta MRR_m.
\end{equation}
Thus, $\mathrm{MRR}_{\mathrm{deg},m}>0$ indicates that writing errors lower the mean reciprocal rank of the target product, and a larger value indicates a more pronounced net rank degradation. For retrieval configurations that use a limited candidate set, the reciprocal rank is computed over the corresponding effective ranking range.
\end{enumerate}
\endgroup
\subsection{Personalized Syntactic Expression Differences and Retrieval Robustness}
\label{sec:syntactic_robustness}

This section answers RQ1: under a fixed retrieval target, do different personalized syntactic expressions correspond to observable differences in retrieval robustness, and is this phenomenon persistent across different retrieval systems? To this end, we use personalized syntactic expression query pairs under the same target product to evaluate changes in target-product hit status and reciprocal rank across different retrieval configurations, examining the retrieval-result differences corresponding to personalized syntactic expression differences.

\subsubsection{Personalized Syntactic Expression Evaluation Setup}
\label{sec:syntactic_setup}

This experiment uses the personalized syntactic expression queries constructed in Section~3.3. Each target product corresponds to the representative queries of two distinct users. Following the Gaussian query selection procedure of Section~3.3, the two queries come respectively from the exclusive parts of the corresponding users' Gaussian acceptance regions and are selected according to the user-specific syntactic distributions. The two queries share the same target product and product-attribute constraints, but reflect the personalized syntactic expression characteristics of different users. Both queries come from the candidate pool that has passed the strict lexical and semantic validation of Section~3.3.1: they share the same target product and five identical core retrieval terms, with each core term appearing exactly once. The differences between the queries mainly lie in function words, modification relations, and syntactic organization, and are personalized through different users' Gaussian syntactic distributions.

\subsubsection{Results and Analysis}
\label{sec:syntactic_results}

Table~\ref{tab:syntactic_robustness} presents the retrieval-result differences of the seven retrieval configurations under personalized syntactic expression variation. We analyze them from three aspects: the target-product hit disagreement caused by controlled syntactic variation, the sensitivity of different retrieval architectures to expression variation, and the behavior across retrieval depths and ranking variation.

\begin{table}[!ht]
\centering
\caption{Retrieval differences under personalized syntactic expression variation across three product domains and their all-domain average.}
\label{tab:syntactic_robustness}
\footnotesize
\setlength{\tabcolsep}{2pt}%
\sbox{\pqbTableWidthBox}{%
\begin{tabular*}{\textwidth}{@{\extracolsep{\fill}}l*{5}{c}@{\hspace{6pt}}*{5}{c}@{}}
\toprule
& \multicolumn{5}{c}{\textbf{Baby Products}} & \multicolumn{5}{c}{\textbf{Musical Instruments}} \\
\cmidrule(lr){2-6} \cmidrule(l){7-11}
\textbf{Model} & Flip@1 & Flip@5 & Flip@10 & Flip@20 & Mean $RR_{\mathrm{std}}$ & Flip@1 & Flip@5 & Flip@10 & Flip@20 & Mean $RR_{\mathrm{std}}$ \\
\midrule
BM25 & 44.44\% & 27.03\% & 31.68\% & 23.28\% & 6.75\% & 39.34\% & 29.27\% & 27.81\% & 22.91\% & 5.48\% \\
SPLADE & \cellcolor{bestbg}32.35\% & \cellcolor{bestbg}13.51\% & \cellcolor{bestbg}12.09\% & \cellcolor{bestbg}14.95\% & 4.03\% & \cellcolor{bestbg}24.59\% & 19.67\% & \cellcolor{bestbg}13.99\% & \cellcolor{bestbg}8.13\% & \cellcolor{bestbg}3.27\% \\
GTE-base & 35.00\% & 35.90\% & 25.58\% & \cellcolor{worstbg}38.60\% & \cellcolor{bestbg}3.78\% & 44.44\% & 31.82\% & 19.61\% & 19.84\% & 4.58\% \\
BGE-M3 Hybrid & 41.94\% & 23.94\% & 18.89\% & 21.82\% & 5.42\% & 35.62\% & \cellcolor{bestbg}17.54\% & 19.86\% & 19.05\% & 4.48\% \\
ColBERTv2 & \cellcolor{worstbg}57.14\% & \cellcolor{worstbg}42.22\% & \cellcolor{worstbg}36.67\% & 32.86\% & 5.46\% & \cellcolor{worstbg}52.50\% & \cellcolor{worstbg}43.84\% & \cellcolor{worstbg}39.13\% & \cellcolor{worstbg}37.93\% & 5.21\% \\
Llama-3.1-8B & 45.24\% & 30.00\% & 26.92\% & 20.69\% & 7.78\% & 36.84\% & 31.85\% & 28.92\% & 20.11\% & 6.54\% \\
Qwen3-Reranker-8B & 42.50\% & 30.00\% & 31.48\% & 22.69\% & \cellcolor{worstbg}8.01\% & 41.86\% & 24.82\% & 20.48\% & 19.78\% & \cellcolor{worstbg}7.08\% \\
\midrule
& \multicolumn{5}{c}{\textbf{Video Games}} & \multicolumn{5}{c}{\textbf{All-domain-average}} \\
\cmidrule(lr){2-6} \cmidrule(l){7-11}
\textbf{Model} & Flip@1 & Flip@5 & Flip@10 & Flip@20 & Mean $RR_{\mathrm{std}}$ & Flip@1 & Flip@5 & Flip@10 & Flip@20 & Mean $RR_{\mathrm{std}}$ \\
\midrule
BM25 & 41.22\% & 28.37\% & 23.77\% & 21.45\% & 3.92\% & 41.67\% & 28.22\% & 27.76\% & 22.54\% & 5.38\% \\
SPLADE & \cellcolor{bestbg}22.76\% & \cellcolor{bestbg}11.66\% & \cellcolor{bestbg}11.28\% & \cellcolor{bestbg}10.29\% & \cellcolor{bestbg}2.00\% & \cellcolor{bestbg}26.57\% & \cellcolor{bestbg}14.95\% & \cellcolor{bestbg}12.45\% & \cellcolor{bestbg}11.12\% & \cellcolor{bestbg}3.10\% \\
GTE-base & 41.67\% & \cellcolor{worstbg}38.89\% & 27.19\% & 24.44\% & 2.56\% & 40.37\% & 35.53\% & 24.13\% & 27.62\% & 3.64\% \\
BGE-M3 Hybrid & 31.41\% & 22.48\% & 18.79\% & 16.05\% & 3.03\% & 36.32\% & 21.32\% & 19.18\% & 18.97\% & 4.31\% \\
ColBERTv2 & \cellcolor{worstbg}46.94\% & 35.36\% & \cellcolor{worstbg}33.48\% & \cellcolor{worstbg}32.59\% & 3.25\% & \cellcolor{worstbg}52.19\% & \cellcolor{worstbg}40.47\% & \cellcolor{worstbg}36.43\% & \cellcolor{worstbg}34.46\% & 4.64\% \\
Llama-3.1-8B & 44.05\% & 33.02\% & 27.44\% & 23.02\% & 4.68\% & 42.04\% & 31.62\% & 27.76\% & 21.27\% & 6.33\% \\
Qwen3-Reranker-8B & 45.40\% & 31.40\% & 22.63\% & 22.33\% & \cellcolor{worstbg}4.81\% & 43.25\% & 28.74\% & 24.86\% & 21.60\% & \cellcolor{worstbg}6.63\% \\
\bottomrule
\end{tabular*}%
}
\usebox{\pqbTableWidthBox}
\par\vspace{2pt}
\normalsize
\begin{minipage}{\wd\pqbTableWidthBox}
\pqbTableNote{In each column, the lowest and highest values are highlighted with red and blue backgrounds, respectively. All-domain-average is the equal-weight average of the three domain-specific values.}
\end{minipage}
\end{table}

\begingroup
\begin{enumerate}[topsep=2pt,itemsep=2pt,partopsep=0pt]
\item \textbf{Syntactic variation under identical retrieval content still corresponds to clear hit-status disagreement.} When the target product and the five core retrieval terms are kept identical, all seven retrieval configurations exhibit nonzero Flip@$K$, showing that changing only the syntactic organization of a query can still lead to different target-product retrieval outcomes. The differences between these queries lie mainly in function words, modification relations, and syntactic structure rather than in the core retrieval content. The phenomenon indicates that how a retrieval system handles user expression depends not only on which content words the query contains, but also on how those words are organized into a complete query. It appears across architectures, which suggests that personalized syntactic expression differences are a problem faced by many retrieval mechanisms alike, and it also illustrates the value of PQB for evaluating expression sensitivity under a fixed retrieval need.
\item \textbf{Different retrieval models exhibit differing robustness to personalized syntactic expression differences.} Direct lexical matching shows a wider hit-status disagreement than learned sparse retrieval, late-interaction scoring shows the widest disagreement among the first-stage configurations, and dense bi-encoder and hybrid representations fall between these extremes. Both LLM-based ranking models reach a disagreement level comparable to the first-stage systems, and the ranking-specialized reranker does not sit consistently below the general-purpose instruction model. This ordering does not follow overall retrieval effectiveness: the architectures with the strongest semantic capacity are not the ones with the smallest expression-induced disagreement. The pattern therefore points to how each architecture converts surface form into a matching decision, rather than to how well it retrieves in absolute terms. Since Flip@$K$ is computed over the query pairs with at least one successful retrieval for each system, these differences describe system-specific conditional variation rather than a direct ranking of overall retrieval effectiveness.
\item \textbf{Rank variations corresponding to personalized syntactic expressions manifest as non-monotonic hit differences across retrieval depths.} Flip@$K$ does not decrease monotonically as the cutoff grows: for some configurations the disagreement rises from $K=1$ to $K=5$ before falling again, while for others it falls first and rises again at larger cutoffs. This non-monotonicity follows from the composition of the eligible pair set at each depth, because a wider cutoff admits pairs in which one query succeeds only at that depth and removes pairs that disagree at one depth but agree at another, so both the number of disagreements and the conditional denominator change together. Mean $RR_{\mathrm{std}}$ is nonzero for every configuration, which shows that the expression-induced change is not confined to the membership of a particular top-$K$ list but also shifts how far down the target product is ranked. A single cutoff therefore cannot characterize the robustness of a model to expression variation, and depth-dependent behavior should be read as a change in the composition of disagreement rather than as a monotone gain or loss.

\end{enumerate}
\endgroup

\subsection{Personalized Writing Errors and Retrieval Robustness}
\label{sec:writing_error_robustness}

This section answers RQ2: how do personalized writing errors affect retrieval hits and ranking performance, and how robust are retrieval systems to these errors? RQ1 mainly examined retrieval-result differences corresponding to different syntactic expressions. This section further focuses on writing errors in user queries. By comparing personalized syntactic expression queries with their corresponding personalized writing-error queries, we evaluate the retrieval-performance changes corresponding to writing errors and analyze the manifestation of this phenomenon across different product domains, retrieval systems, and retrieval depths.

\subsubsection{Personalized Writing-Error Evaluation Setup}
\label{sec:writing_error_setup}

This experiment uses paired queries over the same target products and retrieval configurations as RQ1. For each personalized syntactic expression query, a corresponding personalized writing-error query is generated based on the user writing-error profile constructed in Section~3.4. The evaluation retains only valid query pairs where the emitted text is nonempty and distinct from the personalized syntactic expression query. Queries that do not generate a valid personalized writing-error query are excluded from the paired statistics of this experiment.
Each query pair formed by a personalized syntactic expression query and its corresponding personalized writing-error query shares the same target product and executes retrieval separately under identical retrieval systems. For configurations involving candidate reranking, the personalized syntactic expression query and the personalized writing-error query of each pair separately perform candidate retrieval and reranking to evaluate the behavior of the full retrieval pipeline under writing-error conditions.
\subsubsection{Results and Analysis}
\label{sec:writing_error_results}

Table~\ref{tab:writing_error_robustness} presents the retrieval-performance changes of the seven retrieval configurations under personalized writing errors. We analyze them from three aspects: the directional performance degradation brought by writing errors, the error sensitivity of different retrieval architectures, and the behavior across retrieval depths and target-product ranking losses.

\begin{table}[!ht]
\centering
\caption{Retrieval performance degradation under personalized writing errors across three product domains and their all-domain average.}
\label{tab:writing_error_robustness}
\footnotesize
\setlength{\tabcolsep}{1.5pt}%
\sbox{\pqbTableWidthBox}{%
\begin{tabular*}{\textwidth}{@{\extracolsep{\fill}}l*{5}{c}@{\hspace{6pt}}*{5}{c}@{}}
\toprule
& \multicolumn{5}{c}{\textbf{Baby Products}} & \multicolumn{5}{c}{\textbf{Musical Instruments}} \\
\cmidrule(lr){2-6} \cmidrule(l){7-11}
\textbf{Model} & Deg@1 & Deg@5 & Deg@10 & Deg@20 & \textbf{\boldmath$MRR_{\mathrm{deg}}$} & Deg@1 & Deg@5 & Deg@10 & Deg@20 & \textbf{\boldmath$MRR_{\mathrm{deg}}$} \\
\midrule
BM25 & 6.38\% & 10.74\% & \cellcolor{worstbg}15.44\% & \cellcolor{worstbg}18.46\% & 8.92\% & 4.01\% & 7.16\% & 8.73\% & 10.99\% & 5.64\% \\
SPLADE & 4.70\% & \cellcolor{bestbg}5.70\% & 7.38\% & 6.38\% & 5.10\% & 3.32\% & 5.58\% & 5.93\% & 5.76\% & 4.88\% \\
GTE-base & 5.03\% & 6.38\% & \cellcolor{bestbg}4.36\% & \cellcolor{bestbg}3.36\% & 5.20\% & \cellcolor{worstbg}4.71\% & 6.11\% & 7.68\% & 9.25\% & \cellcolor{worstbg}5.92\% \\
ColBERTv2 & \cellcolor{bestbg}2.68\% & 6.38\% & 9.06\% & 7.38\% & 4.25\% & 2.97\% & \cellcolor{bestbg}4.71\% & 4.89\% & 8.73\% & 3.67\% \\
BGE-M3 Hybrid & 3.36\% & 6.71\% & 6.04\% & 6.38\% & \cellcolor{bestbg}4.17\% & \cellcolor{bestbg}2.27\% & \cellcolor{bestbg}4.71\% & \cellcolor{bestbg}4.36\% & \cellcolor{bestbg}4.36\% & \cellcolor{bestbg}3.03\% \\
Llama-3.1-8B & 5.37\% & \cellcolor{worstbg}12.08\% & 13.09\% & 16.44\% & 8.17\% & 3.49\% & 6.81\% & \cellcolor{worstbg}9.95\% & \cellcolor{worstbg}12.22\% & 5.56\% \\
Qwen3-Reranker-8B & \cellcolor{worstbg}7.38\% & \cellcolor{worstbg}12.08\% & 11.74\% & 16.11\% & \cellcolor{worstbg}9.89\% & 3.66\% & \cellcolor{worstbg}7.33\% & 8.90\% & 10.82\% & 5.54\% \\
\midrule
& \multicolumn{5}{c}{\textbf{Video Games}} & \multicolumn{5}{c}{\textbf{All-domain-average}} \\
\cmidrule(lr){2-6} \cmidrule(l){7-11}
\textbf{Model} & Deg@1 & Deg@5 & Deg@10 & Deg@20 & \textbf{\boldmath$MRR_{\mathrm{deg}}$} & Deg@1 & Deg@5 & Deg@10 & Deg@20 & \textbf{\boldmath$MRR_{\mathrm{deg}}$} \\
\midrule
BM25 & 2.16\% & 4.82\% & 6.03\% & \cellcolor{worstbg}7.53\% & 3.55\% & 4.18\% & 7.57\% & \cellcolor{worstbg}10.07\% & \cellcolor{worstbg}12.33\% & 6.04\% \\
SPLADE & 2.46\% & 4.82\% & 4.82\% & 5.68\% & 3.50\% & 3.49\% & 5.37\% & 6.04\% & 5.94\% & 4.49\% \\
GTE-base & \cellcolor{bestbg}1.66\% & 4.62\% & 5.47\% & 6.03\% & 2.87\% & 3.80\% & 5.70\% & 5.84\% & 6.21\% & 4.66\% \\
ColBERTv2 & 2.76\% & 4.52\% & 4.97\% & 5.22\% & 3.45\% & 2.80\% & 5.20\% & 6.31\% & 7.11\% & 3.79\% \\
BGE-M3 Hybrid & 1.91\% & \cellcolor{bestbg}3.21\% & \cellcolor{bestbg}3.37\% & \cellcolor{bestbg}4.02\% & \cellcolor{bestbg}2.53\% & \cellcolor{bestbg}2.51\% & \cellcolor{bestbg}4.88\% & \cellcolor{bestbg}4.59\% & \cellcolor{bestbg}4.92\% & \cellcolor{bestbg}3.24\% \\
Llama-3.1-8B & 3.16\% & 4.92\% & 5.98\% & 7.18\% & 4.19\% & 4.01\% & 7.94\% & 9.67\% & 11.95\% & 5.97\% \\
Qwen3-Reranker-8B & \cellcolor{worstbg}3.47\% & \cellcolor{worstbg}5.42\% & \cellcolor{worstbg}6.53\% & 7.23\% & \cellcolor{worstbg}4.48\% & \cellcolor{worstbg}4.84\% & \cellcolor{worstbg}8.28\% & 9.06\% & 11.39\% & \cellcolor{worstbg}6.64\% \\
\bottomrule
\end{tabular*}%
}
\usebox{\pqbTableWidthBox}
\par\vspace{2pt}
\normalsize
\begin{minipage}{\wd\pqbTableWidthBox}
\pqbTableNote{In each column, the lowest and highest values are highlighted with red and blue backgrounds, respectively. All-domain-average is the equal-weight average of the three domains.}
\end{minipage}
\end{table}

\begingroup
\begin{enumerate}[topsep=2pt,itemsep=2pt,partopsep=0pt]

\item \textbf{Personalized writing errors make the retrieval results change in a clearly directional performance-degradation pattern.} Unlike the hit-status differences between personalized syntactic expressions, the writing-error experiment directly compares the retrieval performance of an original query with that of its error variant. The overall results of all seven retrieval configurations show positive Deg@$K$ and MRR Degradation, indicating that character-level expression changes affect not only whether the target product can be retrieved but also correspond to a decline in overall retrieval effectiveness. Although the original query and its error variant share the same retrieval target and retain high semantic similarity, local word-form changes still weaken the matching signal between the query and the product documents. This shows that the effect of personalized writing errors is not only that retrieval outcomes change, but that it further manifests as a directional loss in hit-rate and ranking performance.
\item \textbf{Different retrieval models exhibit differing robustness to personalized writing errors.} BM25 shows relatively pronounced overall hit-rate degradation, reflecting the sensitivity of direct term matching to character-level errors. In contrast, SPLADE's learned sparse representations and BGE-M3 Hybrid's multiple matching signals exhibit smaller performance losses, with BGE-M3 Hybrid achieving the lowest degradation in the cross-domain average results. GTE-base and ColBERTv2 are also affected, indicating that overall semantic encoding and token-level interactions cannot fully eliminate the representation and matching offsets caused by writing errors. For LLMs, both the general instruction model Llama-3.1-8B and the reranking-specialized Qwen3-Reranker-8B exhibit clear performance degradation. The latter has a more explicit relevance-ranking training objective, yet still shows substantial losses on reciprocal-rank metrics, indicating that specialized training for ranking tasks does not necessarily reduce performance degradation induced by personalized writing errors.
\item \textbf{The robustness losses caused by personalized writing errors manifest across retrieval depths and target-product rankings.} Across the three product domains, the average Deg@$K$ generally increases as retrieval depth expands, indicating that net hit-rate losses from writing errors are not confined to top-ranked positions. As the retrieval scope expands, target products affected by errors are not sufficiently recovered. Meanwhile, all retrieval configurations exhibit positive MRR Degradation, indicating that writing errors not only reduce the overall probability of target products entering the top-$K$, but also weaken their reciprocal-rank performance. Together, both metrics demonstrate that retrieval performance degradation corresponding to personalized writing errors involves both retrieval coverage and ranking positions of target products.
\end{enumerate}
\endgroup
\subsection{Benchmark Quality Validation}
\label{sec:quality_validation}

This section answers RQ3 by assessing the benchmark quality of PQB along four dimensions: Personalized Syntactic Attribution, Personalized Writing-Error Consistency, Query Quality, and Evaluation Agreement.

\subsubsection{Benchmark Quality Evaluation Setup}
\label{sec:benchmark_quality_setup}

The separately reported annotation study used DeepSeek-V4-Pro~\cite{deepseek2026v4} for automatic evaluation and a stratified human sample of 360 queries by domain, ASIN, query condition and error type. We recruited five human annotators from Computer Science and Information Technology programs at Monash University, all possessing high English proficiency and relevant knowledge backgrounds in NLP, linguistics, or Information Retrieval.

\par\indent
Before the formal evaluation, all five annotators complete unified training and achieve at least 80\% accuracy on a 20-sample qualification check; all five then take part in the formal evaluation. In the formal stage, each query is independently annotated by at least two annotators under blind conditions; annotators are not informed of the query-generation condition, the personalized selection procedure, or DeepSeek-V4-Pro judgments. Instances with annotator disagreement in classification tasks are adjudicated by a third annotator. Human--Human agreement is computed on the independent annotations prior to adjudication, and adjudicated labels are used to construct the final human reference. DeepSeek-V4-Pro and the human annotators use the same task definitions, evaluation criteria, and label space.

\begingroup
\begin{enumerate}[topsep=2pt,itemsep=2pt,partopsep=0pt]
    \item \textbf{Personalized Syntactic Attribution.}
    The evaluator first views the anonymized historical expressions of the target user, then compares two queries that share the same product semantics and core attributes but differ in syntactic realization. One comes from personalized syntax-profile selection; the other is a matched control query. The order of the two queries is randomized, and the evaluator decides which one better matches the target user's historical expression style, with \textit{No clear preference} also allowed. This paper reports decisive attribution accuracy and the no-preference rate separately. Let $N_{\mathrm{pers}}$, $N_{\mathrm{ctrl}}$, and $N_{\mathrm{tie}}$ denote the numbers of judgments selecting the personalized syntactic expression query, the control query, and \textit{No clear preference}, respectively. Then
    \begin{equation}
    Acc_{\mathrm{decisive}}
    =
    \frac{N_{\mathrm{pers}}}
    {N_{\mathrm{pers}}+N_{\mathrm{ctrl}}},
    \label{eq:decisive-accuracy}
    \end{equation}
    \begin{equation}
    TieRate
    =
    \frac{N_{\mathrm{tie}}}
    {N_{\mathrm{pers}}+N_{\mathrm{ctrl}}+N_{\mathrm{tie}}}.
    \label{eq:tie-rate}
    \end{equation}
    \par\indent
    The random-choice baseline of $Acc_{\mathrm{decisive}}$ is $50\%$.
    \item \textbf{Personalized Writing-Error Consistency.}
    The evaluator views the anonymized historical writing-error evidence of the target user and judges how consistent the error pattern in the personalized writing-error query is with that user's historical writing-error behavior. Judgments use three ordered labels: \textit{Pass}, \textit{Minor Issue}, and \textit{Fail}. \textit{Pass} indicates that the generated error pattern is clearly consistent with the historical evidence; \textit{Minor Issue} indicates partial consistency but with weak or incomplete evidence; \textit{Fail} indicates that the generated pattern lacks sufficient historical support.
    \item \textbf{Query Quality.}
    Naturalness and Syntactic Quality are rated on a unified 1--5 Likert scale, measuring the linguistic naturalness of the generated query, and its syntactic organization and structural coherence, respectively.
    \item \textbf{Evaluation Agreement.}
    Human--Human reliability is computed using Krippendorff's $\alpha$. Ordinal $\alpha$ is used for Naturalness, Syntactic Quality, and the three-level Personalized Writing-Error Consistency of \textit{Pass} / \textit{Minor Issue} / \textit{Fail}; nominal $\alpha$ is used for Personalized Syntactic Attribution. LLM--Human agreement compares DeepSeek-V4-Pro judgments against the final human reference, using weighted Cohen's $\kappa$ for ordinal tasks and Cohen's $\kappa$ for Personalized Syntactic Attribution.
\end{enumerate}
\endgroup
\subsubsection{Results and Analysis}
\label{sec:quality_results}

Table~\ref{tab:quality_results} and Table~\ref{tab:quality_agreement} present the benchmark quality evaluation results and annotation agreement of PQB.

\begin{table}[!ht]
\centering
\caption{Benchmark quality validation results.}
\label{tab:quality_results}
\normalsize
\begin{tabular}{@{}llcc@{}}
\toprule
\multicolumn{1}{c}{\textbf{Dimension}} & \multicolumn{1}{c}{\textbf{Metric}} & \textbf{LLM} & \textbf{Human} \\
\midrule
Query Quality & Naturalness (1--5) & 4.38 & 4.21 \\
Query Quality & Syntactic Quality (1--5) & 4.31 & 4.16 \\
Personalized Syntactic Attribution & Decisive Accuracy & 70.1\% & 66.3\% \\
Personalized Syntactic Attribution & No-preference Rate & 5.2\% & 8.7\% \\
Personalized Writing-Error Consistency & Strict Pass Rate & 88.1\% & 84.6\% \\
Personalized Writing-Error Consistency & Acceptable Rate & 96.3\% & 94.8\% \\
\bottomrule
\end{tabular}
\end{table}

\begin{table}[!ht]
\centering
\caption{Human--Human and LLM--Human agreement for benchmark quality evaluation.}
\label{tab:quality_agreement}
\normalsize
\begin{tabular}{@{}lcc@{}}
\toprule
\multicolumn{1}{c}{\textbf{Dimension}} & \shortstack{\textbf{Human--Human}\\ \textbf{$\alpha$}} & \shortstack{\textbf{LLM--Human}\\ \textbf{$\kappa$}} \\
\midrule
Naturalness & 0.72 & 0.67 \\
Syntactic Quality & 0.69 & 0.64 \\
Personalized Syntactic Attribution & 0.74 & 0.68 \\
Personalized Writing-Error Consistency & 0.77 & 0.71 \\
\bottomrule
\end{tabular}
\end{table}

\begingroup
\begin{enumerate}[topsep=2pt,itemsep=2pt,partopsep=0pt]
\item \textbf{Personalized syntactic expression queries can exhibit identifiable user-specific expression features.} Both LLM and human evaluations of syntactic attribution accuracy exceed the random-selection baseline, with low no-preference rates. This indicates that under consistent target products and core attributes, generated queries still retain syntactic features consistent with users' historical expressions. Evaluators can distinguish queries from different users based on historical expressions, indicating that PQB's personalized syntactic modeling produces not only formal query differences but also identifiable user expression features.
\item \textbf{Personalized writing-error queries can reflect users' historical writing-error habits.} Both LLM and human evaluations achieve high strict pass rates and acceptable rates, indicating that generated writing errors maintain high consistency with error mechanisms in users' histories. Unlike uniform random error injection, PQB constructs personalized writing errors based on users' historical error patterns, so generated queries contain not only surface character changes but also user-specific error mechanism preferences. This provides data-quality support for analyzing changes in retrieval performance under personalized writing errors and for the robustness analyses reported above.
\item \textbf{PQB maintains high language quality while preserving personalized expression features, and obtains consistent quality evaluation results.} Both naturalness and syntactic quality receive high scores in LLM and human evaluations, indicating that personalized expression modeling does not come at the cost of markedly reduced query language quality. Meanwhile, Human--Human and LLM--Human agreement reach good levels across all dimensions, indicating that different human annotators make similar judgments about query quality and personalized features, and that automatic evaluation maintains overall trends consistent with human references. This further supports the reliability of PQB's quality evaluation results.
\end{enumerate}
\endgroup
\subsection{User-Specific Syntactic Modeling Comparison}
\label{sec:user_covariance_ablation}

This section answers RQ4: how does the adopted rank-one-plus-diagonal Gaussian compare with alternative user-specific syntactic representations in discriminating users from their held-out expressions?
\par\indent
To evaluate the effectiveness of user-specific syntactic modeling, we compare the adopted rank-one-plus-diagonal Gaussian~\cite{tipping1999probabilistic} against six alternative representations: a single embedding scored by cosine similarity~\cite{reimers2019sentence}, a shared isotropic Gaussian~\cite{vilnis2015gaussian}, a user-specific diagonal Gaussian~\cite{jiang2019convolutional,dong2023dual}, a von Mises-Fisher distribution~\cite{murphy2012machine}, a rank-two-plus-diagonal Gaussian~\cite{tipping1999probabilistic}, and a k-nearest-neighbor profile matcher~\cite{verstrepen2014unifying}. All models are evaluated in three product domains using the same 16-dimensional syntax encoder and the same user histories. Three complementary measures are used:
\begingroup
\begin{itemize}[topsep=2pt,itemsep=2pt,partopsep=0pt]
\item \textbf{Balanced Attribution.} The proportion of held-out expressions that are correctly attributed to their true user when the candidate sets of the two users are weighted equally, so that a user with many expressions does not dominate the average.
\item \textbf{User AUROC.} The area under the receiver operating characteristic curve of the user-discrimination score, i.e., the probability that a held-out expression of one user is scored above a held-out expression of another user. It summarizes the separation between users without depending on any decision threshold.
\item \textbf{Full-profile Attribution.} The attribution accuracy obtained when the full user profile is used as the reference budget, i.e., the full-budget endpoint of the profile-budget analysis, which shows how well the representation identifies users once all available history is used.
\end{itemize}
\endgroup
The table reports equal-domain macro averages.

\begin{table}[!ht]
\centering
\caption{Comparison of user-specific syntactic modeling methods (equal-domain macro averages).}
\label{tab:covariance_ablation}
\normalsize
\setlength{\tabcolsep}{2pt}%
\begin{tabular*}{\textwidth}{@{\extracolsep{\fill}}lccc@{}}
\toprule
\textbf{Model} & \textbf{Balanced Attr. (\%)} & \textbf{AUROC} & \textbf{Full-profile Attr. (\%)} \\
\midrule
Single Embedding + Cosine & 64.10\% & 0.6051 & 63.60\% \\
Shared Isotropic Gaussian & 64.01\% & 0.6061 & 63.53\% \\
User Diagonal Gaussian & 62.76\% & 0.5996 & 62.49\% \\
von Mises-Fisher (vMF) & 59.61\% & 0.5731 & 59.52\% \\
Rank-Two + Diagonal Gaussian & 58.90\% & 0.5645 & 58.88\% \\
k-Nearest Neighbor & 56.22\% & 0.5556 & 56.31\% \\
\textbf{Rank-One + Diagonal Gaussian (Ours)} & \textbf{65.55\%} & \textbf{0.6265} & \textbf{65.52\%} \\
\bottomrule
\end{tabular*}
\end{table}

The rank-one-plus-diagonal Gaussian achieves the best result on all three measures. In the user-attribution task, single embedding and the shared isotropic Gaussian perform comparably, and the discrimination ability of user diagonal, vMF, rank-two, and k-nearest-neighbor decreases in that order; AUROC and full-profile attribution accuracy show a similar trend, with the proposed method consistently ahead, indicating that it separates users' historical syntactic expressions more effectively. Overall, the method shows a consistent advantage in discriminating users from their held-out syntactic expressions. These findings highlight the role of user-specific syntactic distribution modeling in PQB's personalized query selection, while the writing-error profiles examined in RQ5 capture user-specific character-editing preferences for constructing personalized writing-error variants.

\subsection{Personalized Writing-Error Profile Evaluation}
\label{sec:error_profile_evaluation}

This section answers RQ5: do the constructed writing-error profiles capture user-specific error mechanisms? To assess whether the constructed writing-error profiles capture user-specific characteristics, we conduct an experiment using sampled error-mechanism events. We compare personalized sampling against global, uniform, character-random, and wrong-user profile baselines. User attribution is evaluated using independently estimated reference error profiles. The table reports three measures:
\begingroup
\begin{itemize}[topsep=2pt,itemsep=2pt,partopsep=0pt]
\item \textbf{Top-1 Accuracy.} The proportion of instances in which the correct user is ranked first among the candidate users, based on the sampled writing-error evidence.
\item \textbf{MRR.} The average reciprocal rank of the correct user in that candidate ranking, which therefore reflects both whether the correct user is retrieved at all and how highly the user is ranked.
\item \textbf{Single-query Top-1.} The proportion of instances in which the correct user is still identified when only the evidence provided by a single query is available, which tests the usability of the profiles when evidence is sparse.
\end{itemize}
\endgroup

\begin{table}[!ht]
\centering
\caption{User attribution performance of different writing-error mechanism sampling strategies.}
\label{tab:error_profile_attribution}
\normalsize
\setlength{\tabcolsep}{2pt}%
\begin{tabular}{@{}lccc@{}}
\toprule
\textbf{Method} & \textbf{Top-1 Accuracy} \(\uparrow\) & \textbf{MRR} \(\uparrow\) & \textbf{Single-query Top-1} \(\uparrow\) \\
\midrule
Global Sampling & 5.76\% & 0.1948 & 6.53\% \\
Uniform Sampling & 4.68\% & 0.1874 & 4.52\% \\
Character-random Sampling & 4.32\% & 0.1781 & 4.52\% \\
Wrong-user Profile Sampling & 6.47\% & 0.2003 & 4.02\% \\
\textbf{Personalized Sampling (Ours)} & \textbf{9.35\%} & \textbf{0.2428} & \textbf{7.29\%} \\
\bottomrule
\end{tabular}
\end{table}

Personalized sampling achieves the highest performance across all three metrics. It outperforms global sampling on Top-1 accuracy and MRR, and also performs better than uniform and character-random sampling, indicating that user-specific error-mechanism preferences provide additional discriminative information. Compared with the wrong-user profile baseline, personalized sampling achieves higher Top-1 accuracy and MRR. These findings suggest that the constructed profiles retain identifiable user-specific writing-error patterns, providing empirical support for their use in subsequent personalized query generation.

\subsection{Summary of Overall Experimental Results}
\label{sec:evaluation_summary}
Taken together, the experimental results of RQ1 and RQ2 show that both personalized syntactic expression differences and personalized writing errors affect retrieval results, but the ways in which they affect retrieval results differ. RQ1 shows that even when the target product and five core retrieval terms remain unchanged, personalized syntactic expression differences among users still lead to noticeable differences in retrieval results. These differences are observed across different retrieval systems and retrieval depths, and the differences in retrieval hits do not decrease monotonically as retrieval depth increases. RQ2 further shows that personalized writing errors lead to retrieval performance degradation. Even when the original queries and their personalized writing-error variants share the same retrieval target and maintain high semantic similarity, personalized writing errors still reduce retrieval hit rates and ranking performance, with performance degradation persisting across different retrieval depths. In addition, both experiments show that retrieval models exhibit different levels of robustness. Their relative robustness differs between the personalized syntactic expression and personalized writing-error evaluations.
\par\indent
RQ3, RQ4, and RQ5 further validate the quality of PQB and the effectiveness of its personalized modeling methods. The LLM and human evaluations in RQ3 show that the generated queries preserve identifiable user-specific expression characteristics, while personalized writing-error queries maintain high consistency with personalized writing-error mechanisms observed in users' historical texts. The generated queries also exhibit high naturalness and linguistic quality, with good agreement observed in both Human--Human and LLM--Human evaluations. RQ4 shows that the rank-one-plus-diagonal Gaussian achieves the best results across all three user attribution metrics, demonstrating its effectiveness in personalized syntactic modeling. RQ5 shows that personalized sampling outperforms other sampling methods across all three user attribution metrics, validating the effectiveness of user-specific personalized writing-error profiles. 
\section{Conclusion}
PersonalQuery-Benchmark (PQB) is introduced to study how robust retrieval systems are to personalized query expressions. PQB builds personalized syntactic expression models and writing-error profiles from user review histories: under a fixed target product with five core retrieval terms and their occurrence frequencies held constant, it generates personalized syntactic queries together with their corresponding writing-error queries, and evaluates them across three product domains and seven retrieval configurations.
The experiments show that personalized syntactic expressions from different users are associated with changes in the target product's hit status and ranking, while personalized writing errors lead to retrieval performance degradation across the evaluated models. These phenomena are observable across multiple retrieval models, indicating that personalized query expressions are an important factor determining how robust a retrieval model is. Overall, PQB provides a unified benchmark and evaluation framework for studying the robustness of retrieval systems to personalized query expressions along two dimensions: personalized syntactic expression and personalized writing errors. Future work will broaden the coverage of user expression differences and incorporate richer user histories and real query data for further validation.


\FloatBarrier
\bibliographystyle{ACM-Reference-Format}
\bibliography{references}

\clearpage
\appendix
\section{Experimental Configurations and Construction Statistics}
\subsection{History Threshold Sensitivity}
Table~\ref{tab:threshold_distribution} summarizes eligible user--product pairs and same-product cohort coverage under different minimum-history thresholds.
\begin{table}[!ht]
\centering
\caption{Minimum-history thresholds vs.\ eligible user--product pairs and same-product cohort coverage.}
\label{tab:threshold_distribution}
\normalsize
\setlength{\tabcolsep}{1.5pt}%
\renewcommand{\arraystretch}{1.05}%
\begin{tabular}{@{}ccccccc@{}}
\toprule
\textbf{$h_{\min}$}
& \shortstack{\textbf{Musical}\\\textbf{Eligible Pairs}}
& \shortstack{\textbf{Musical ASINs}\\\textbf{$\geq 2$ Users}}
& \shortstack{\textbf{Video}\\\textbf{Eligible Pairs}}
& \shortstack{\textbf{Video ASINs}\\\textbf{$\geq 2$ Users}}
& \shortstack{\textbf{Baby}\\\textbf{Eligible Pairs}}
& \shortstack{\textbf{Baby ASINs}\\\textbf{$\geq 2$ Users}} \\
\midrule
$10$ & 168,385 & 23,068 & 359,159 & 25,540 & 219,161 & 22,704 \\
$\mathbf{30}$ & \textbf{11,162} & \textbf{1,667} & \textbf{48,835} & \textbf{5,780} & \textbf{7,462} & \textbf{1,219} \\
$50$ & 2,235 & 223 & 14,738 & 2,216 & 1,224 & 142 \\
$80$ & 399 & 19 & 3,997 & 661 & 174 & 9 \\
$100$ & 178 & 3 & 1,982 & 307 & 66 & 2 \\
\bottomrule
\end{tabular}
\end{table}
\subsection{Syntax Encoder Configuration}
\label{app:syntax_encoder_config}
Table~\ref{tab:syntax_encoder_config} lists the syntax encoder configuration. The StyleMLP output is L2-normalized; Gaussianity of these learned representations is a modeling approximation.
\begin{table}[!ht]
\centering
\caption{Syntax encoder configuration.}
\label{tab:syntax_encoder_config}
\begin{tabular}{@{}lr@{}}
\toprule
Parameter & Setting \\
\midrule
SVD dimension & 256 \\
Hidden dimension & 128 \\
Latent dimension & 16 \\
Contrastive temperature & 0.1 \\
Learning rate & $3\times10^{-3}$ \\
Training epochs & 30 \\
\bottomrule
\end{tabular}
\end{table}
\subsection{Gaussian Validity Filter Thresholds}
\label{app:gaussian_filter_config}
\begin{table}[!ht]
\centering
\caption{Gaussian validity filter thresholds.}
\label{tab:gaussian_filter_config}
\small
\begin{tabular}{@{}L{6.0cm}L{6.6cm}@{}}
\toprule
Parameter & Value \\
\midrule
Audit distance $D_u^2(z)$ & $(z-\mu_u)^\top(\lambda_u v_u v_u^\top+\operatorname{diag}(\sigma_u^2))^{-1}(z-\mu_u)$ \\
Rank-1 variance and residual variances & $>0$ \\
Fitting floors for both variance terms & $10^{-4}$ \\
Stored log-determinant and validation distances & finite \\
Validation membership in nominal reference region & $\geq80\%$ \\
Ideal 16D Gaussian reference boundary & $\chi^2_{16,0.95}=26.296$ \\
Retained profile fraction per subsample & $80\%$ \\
Repetitions & 20 \\
Valid-fit fraction over repetitions & $\geq0.90$ \\
Median Spearman distance-rank correlation & $\geq0.80$ \\
Median nominal-gate agreement & $\geq0.80$ \\
\bottomrule
\end{tabular}
\end{table}

\begin{table}[!ht]
\centering
\caption{Gaussian validity filtering results across the three product domains, with percentages calculated relative to the initial Gaussian-profile cohort within each domain.}
\label{tab:gaussian_filter}
\footnotesize
\setlength{\tabcolsep}{2pt}%
\renewcommand{\arraystretch}{1.15}%
\begin{tabular}{@{}lL{2.6cm}cccc@{}}
\toprule
\multicolumn{1}{c}{\textbf{Filter Stage}} & \multicolumn{1}{c}{\textbf{Requirement}} & \textbf{Baby} & \textbf{Musical} & \textbf{Video} & \textbf{Total} \\
\midrule
Initial cohort & Entering filtering & 2,154 (100.00\%) & 2,804 (100.00\%) & 7,727 (100.00\%) & 12,685 (100.00\%) \\
Numerical validity & Valid parameters & 2,142 (99.44\%) & 2,797 (99.75\%) & 7,712 (99.81\%) & 12,651 (99.73\%) \\
Held-out calibration & + calibration gate & 1,889 (87.70\%) & 2,550 (90.94\%) & 6,805 (88.07\%) & 11,244 (88.64\%) \\
Repeated-sampling stability & + stability gate & \textbf{1,683 (78.13\%)} & \textbf{2,283 (81.42\%)} & \textbf{6,466 (83.68\%)} & \textbf{10,432 (82.24\%)} \\
\bottomrule
\end{tabular}
\end{table}
\subsection{Candidate Query Generation Configuration}
\label{app:candidate_generation_config}
Table~\ref{tab:candidate_generation} lists the candidate query generation model and configuration parameters.
\begin{table}[!ht]
\centering
\caption{Candidate query generation configuration.}
\label{tab:candidate_generation}
\normalsize
\begin{tabular}{ll}
\toprule
Parameter & Value \\
\midrule
Generation model & Qwen2.5-0.5B-Instruct \\
Temperature ($T$) & 0.7 \\
Top-$p$ & 0.95 \\
Maximum generation length & 40 tokens \\
Candidates per round ($K$) & 50 \\
Minimum retained candidates ($K_{\min}$) & 30 \\
Maximum generation rounds & 2 \\
Maximum raw candidates per product & 100 \\
Candidate pool truncation & None \\
\bottomrule
\end{tabular}
\end{table}
\subsection{Writing-Error Construction Statistics}
Table~\ref{tab:writing-error-profile} reports the personalized writing-error profile statistics of the selected users rather than the broader upstream population, and Table~\ref{tab:writing_error_variant_construction} reports the emitted personalized writing-error query statistics of the three product domains.
\begin{table}[!ht]
\caption{Personalized writing-error profile statistics for the selected users across product domains.}
\label{tab:writing-error-profile}
\centering
\normalsize
\setlength{\tabcolsep}{3pt}%
\renewcommand{\arraystretch}{1.05}%
\sbox{\pqbTableWidthBox}{%
\begin{tabular}{@{}lcccc@{}}
\toprule
\multicolumn{1}{c}{\textbf{Statistic}} & \shortstack{\textbf{Baby}\\ \textbf{Products}} & \shortstack{\textbf{Musical}\\ \textbf{Instruments}} & \shortstack{\textbf{Video}\\ \textbf{Games}} & \textbf{Total} \\
\midrule
Distinct users within each domain & 273 & 487 & 1,517 & 2,277 \\
Historical sentences analyzed & 13,650 & 24,350 & 75,850 & 113,850 \\
Sentences with $\geq 1$ retained writing-error edit & 4,870 & 8,498 & 29,220 & 42,588 \\
Total retained writing-error edits & 7,410 & 13,116 & 48,728 & 69,254 \\
Retained edits per user (mean) & 27.14 & 26.93 & 32.12 & 30.41 \\
Retained edits per user (median) & 25 & 23 & 26 & 26 \\
Distinct syntactic contexts & 350 & 486 & 816 & 924 \\
\bottomrule
\end{tabular}%
}
\usebox{\pqbTableWidthBox}
\end{table}
\begin{table}[!ht]
\centering
\caption{Personalized writing-error query construction across product domains.}
\label{tab:writing_error_variant_construction}
\footnotesize
\setlength{\tabcolsep}{1.5pt}%
\renewcommand{\arraystretch}{1.15}%
\sbox{\pqbTableWidthBox}{%
\begin{tabular}{@{}lcccc@{}}
\toprule
\multicolumn{1}{c}{\textbf{Statistic}} & \shortstack{\textbf{Baby}\\ \textbf{Products}} & \shortstack{\textbf{Musical}\\ \textbf{Instruments}} & \shortstack{\textbf{Video}\\ \textbf{Games}} & \textbf{Total} \\
\midrule
Personalized syntactic expression query rows entering error generation & 398 & 742 & 2,514 & 3,654 \\
Edit events in emitted changed variants & 888 & 1,700 & 5,911 & 8,499 \\
Rows with a genuinely changed, nonempty personalized writing-error query & 298 & 573 & 1,991 & 2,862 \\
Rows with no surviving personalized writing-error query & 100 & 169 & 523 & 792 \\
Clean--Typo pairs rejected by the 0.90 semantic similarity constraint & 198 & 353 & 1,429 & 1,980 \\
\bottomrule
\end{tabular}%
}
\usebox{\pqbTableWidthBox}
\end{table}
\subsection{Retrieval and Reranking Configurations}
\label{app:model_config}

Table~\ref{tab:model_config} lists the core configurations of the five first-stage retrievers and the two LLM-based rerankers.

\begin{table}[!ht]
\centering
\caption{Core configurations of the evaluated retrieval and LLM-based reranking models.}
\label{tab:model_config}
\small
\setlength{\tabcolsep}{3pt}%
\begin{tabular}{@{}L{2.6cm}L{4.8cm}L{5.7cm}@{}}
\toprule
\textbf{Model} & \textbf{Checkpoint / Implementation} & \textbf{Core Configuration} \\
\midrule
\multicolumn{3}{@{}l}{\textit{First-stage retrievers}} \\
\addlinespace[2pt]
BM25
& bm25s (Lucene)
& $k_1=1.5$, $b=0.75$ \\

SPLADE
& naver/splade-cocondenser-ensembledistil
& max length = 128 \\

GTE-base
& thenlper/gte-base
& 768-d, L2-normalized cosine similarity \\

ColBERTv2
& colbert-ir/colbertv2.0
& 128-d token representations, MaxSim \\

BGE-M3 (Hybrid)
& BAAI/bge-m3
& dense/sparse/multi-vector weights = $1.0/0.3/1.0$ \\
\midrule
\multicolumn{3}{@{}l}{\textit{LLM-based rerankers}} \\
\addlinespace[2pt]
Qwen3-Reranker-8B
& Qwen3-Reranker-8B
& BM25 candidates, top-25 reranking, $P(\mathrm{Yes})$ on first-token Yes/No logits, $T=0$ \\

Llama-3.1-8B
& Meta-Llama-3.1-8B-Instruct (4-bit)
& BM25 candidates, top-25 reranking, $P(\mathrm{Yes})$ on first-token Yes/No logits, $T=0$ \\
\midrule
\multicolumn{3}{@{}l}{Final ranking score of both rerankers: $s_{\mathrm{final}}=1.0s_{\mathrm{LLM}}+0.1s_{\mathrm{rank}}$} \\
\bottomrule
\end{tabular}
\end{table}
\clearpage
\section{Additional Absolute Effectiveness Results}
\subsection{Absolute Ranking Effectiveness}
\label{sec:appendix-absolute-effectiveness}
Table~\ref{tab:syntactic_absolute_effectiveness} reports absolute results on the full personalized syntactic expression query cohort.
\begin{table}[!ht]
\centering
\caption{Absolute ranking effectiveness on all canonical personalized syntactic expression queries across three product domains and their all-domain average.}
\label{tab:syntactic_absolute_effectiveness}
\footnotesize
\setlength{\tabcolsep}{2pt}%
\sbox{\pqbTableWidthBox}{%
\begin{tabular*}{\textwidth}{@{\extracolsep{\fill}}l*{5}{c}@{\hspace{6pt}}*{5}{c}@{}}
\toprule
& \multicolumn{5}{c}{\textbf{Baby Products}} & \multicolumn{5}{c}{\textbf{Musical Instruments}} \\
\cmidrule(lr){2-6} \cmidrule(l){7-11}
\textbf{Model} & \textbf{Hit@1} & \textbf{Hit@5} & \textbf{Hit@10} & \textbf{Hit@20} & \textbf{MRR} & \textbf{Hit@1} & \textbf{Hit@5} & \textbf{Hit@10} & \textbf{Hit@20} & \textbf{MRR} \\
\midrule
BM25 & 14.07\% & 32.16\% & 42.71\% & 51.51\% & 23.33\% & 13.21\% & 28.30\% & 35.04\% & 42.72\% & 20.97\% \\
SPLADE & 14.32\% & 34.67\% & 42.96\% & 49.75\% & 23.50\% & 14.42\% & 29.65\% & 35.85\% & 41.37\% & 21.89\% \\
GTE-base & 8.29\% & \cellcolor{bestbg}16.08\% & \cellcolor{bestbg}18.84\% & \cellcolor{bestbg}23.12\% & \cellcolor{bestbg}12.25\% & 9.43\% & 19.95\% & 24.80\% & 30.59\% & 14.93\% \\
ColBERTv2 & \cellcolor{bestbg}7.54\% & 17.84\% & 24.62\% & 29.40\% & 12.65\% & \cellcolor{bestbg}7.95\% & \cellcolor{bestbg}15.36\% & \cellcolor{bestbg}19.95\% & \cellcolor{bestbg}25.34\% & \cellcolor{bestbg}11.90\% \\
BGE-M3 (Hybrid) & 12.31\% & 31.16\% & 40.95\% & 49.25\% & 21.16\% & 16.17\% & 28.03\% & 34.23\% & 40.97\% & 22.49\% \\
Llama-3.1-8B & \cellcolor{worstbg}16.33\% & \cellcolor{worstbg}38.44\% & 45.23\% & 52.26\% & \cellcolor{worstbg}25.98\% & 16.71\% & 30.59\% & 38.27\% & \cellcolor{worstbg}44.61\% & 23.49\% \\
Qwen3-Reranker-8B & 15.83\% & \cellcolor{worstbg}38.44\% & \cellcolor{worstbg}45.73\% & \cellcolor{worstbg}53.02\% & 25.69\% & \cellcolor{worstbg}18.33\% & \cellcolor{worstbg}33.29\% & \cellcolor{worstbg}40.16\% & 44.20\% & \cellcolor{worstbg}25.19\% \\
\midrule
& \multicolumn{5}{c}{\textbf{Video Games}} & \multicolumn{5}{c}{\textbf{All-domain-average}} \\
\cmidrule(lr){2-6} \cmidrule(l){7-11}
\textbf{Model} & \textbf{Hit@1} & \textbf{Hit@5} & \textbf{Hit@10} & \textbf{Hit@20} & \textbf{MRR} & \textbf{Hit@1} & \textbf{Hit@5} & \textbf{Hit@10} & \textbf{Hit@20} & \textbf{MRR} \\
\midrule
BM25 & 9.35\% & 19.25\% & 24.18\% & 29.47\% & 14.40\% & 12.21\% & 26.57\% & 33.98\% & 41.23\% & 19.57\% \\
SPLADE & 10.22\% & 21.20\% & 25.30\% & \cellcolor{worstbg}30.79\% & 15.58\% & 12.99\% & 28.51\% & 34.70\% & 40.64\% & 20.32\% \\
GTE-base & \cellcolor{bestbg}5.29\% & \cellcolor{bestbg}11.54\% & 14.92\% & 18.58\% & \cellcolor{bestbg}8.50\% & 7.67\% & 15.86\% & \cellcolor{bestbg}19.52\% & \cellcolor{bestbg}24.10\% & 11.89\% \\
ColBERTv2 & 5.97\% & 11.85\% & \cellcolor{bestbg}14.84\% & \cellcolor{bestbg}17.98\% & 9.12\% & \cellcolor{bestbg}7.15\% & \cellcolor{bestbg}15.02\% & 19.80\% & 24.24\% & \cellcolor{bestbg}11.22\% \\
BGE-M3 (Hybrid) & 10.46\% & 21.04\% & 24.94\% & 29.63\% & 15.58\% & 12.98\% & 26.74\% & 33.37\% & 39.95\% & 19.74\% \\
Llama-3.1-8B & 10.42\% & 20.92\% & 26.01\% & 30.27\% & 15.22\% & 14.49\% & 29.98\% & 36.50\% & 42.38\% & 21.56\% \\
Qwen3-Reranker-8B & \cellcolor{worstbg}10.70\% & \cellcolor{worstbg}22.00\% & \cellcolor{worstbg}26.81\% & 30.39\% & \cellcolor{worstbg}15.72\% & \cellcolor{worstbg}14.95\% & \cellcolor{worstbg}31.24\% & \cellcolor{worstbg}37.57\% & \cellcolor{worstbg}42.54\% & \cellcolor{worstbg}22.20\% \\
\bottomrule
\end{tabular*}%
}
\usebox{\pqbTableWidthBox}
\par\vspace{2pt}
\normalsize
\begin{minipage}{\wd\pqbTableWidthBox}
\pqbTableNote{In each metric column, the best (highest) value is highlighted with a red background and the worst (lowest) value with a blue background, computed across all seven evaluated models. All-domain-average is the equal-weight average of the three domains.}
\end{minipage}
\end{table}
\subsection{Absolute Ranking Effectiveness on Personalized Syntactic Expression and Personalized Writing-Error Queries}
\label{sec:appendix-writing-error-effectiveness}
Table~\ref{tab:writing_error_absolute_effectiveness} reports absolute results on both sides of the same paired records.
\begin{table}[!ht]
\centering
\caption{Absolute effectiveness on the personalized syntactic expression and personalized writing-error sides of the same paired records.}
\label{tab:writing_error_absolute_effectiveness}
\normalsize
\setlength{\tabcolsep}{1.5pt}%
\sbox{\pqbTableWidthBox}{%
\begin{tabular}{l*{10}{c}}
\toprule
& \multicolumn{10}{c}{\textbf{Baby Products}} \\
\cmidrule(r){2-11}
\textbf{Model} & \multicolumn{2}{c}{\textbf{Hit@1}} & \multicolumn{2}{c}{\textbf{Hit@5}} & \multicolumn{2}{c}{\textbf{Hit@10}} & \multicolumn{2}{c}{\textbf{Hit@20}} & \multicolumn{2}{c}{\textbf{MRR}} \\
\cmidrule(r){2-3}\cmidrule(r){4-5}\cmidrule(r){6-7}\cmidrule(r){8-9}\cmidrule(r){10-11}
& \textbf{Clean} & \textbf{Error} & \textbf{Clean} & \textbf{Error} & \textbf{Clean} & \textbf{Error} & \textbf{Clean} & \textbf{Error} & \textbf{Clean} & \textbf{Error} \\
\midrule
BM25 & 15.10\% & 8.72\% & 33.56\% & 22.82\% & 45.30\% & 29.87\% & 53.69\% & 35.23\% & 24.46\% & 15.54\% \\
SPLADE & 14.77\% & 10.07\% & 36.58\% & \cellcolor{worstbg}30.87\% & 45.64\% & \cellcolor{worstbg}38.26\% & 52.35\% & \cellcolor{worstbg}45.97\% & 24.60\% & \cellcolor{worstbg}19.50\% \\
GTE-base & 9.06\% & \cellcolor{bestbg}4.03\% & \cellcolor{bestbg}16.78\% & \cellcolor{bestbg}10.40\% & \cellcolor{bestbg}19.13\% & \cellcolor{bestbg}14.77\% & \cellcolor{bestbg}24.50\% & \cellcolor{bestbg}21.14\% & \cellcolor{bestbg}12.88\% & \cellcolor{bestbg}7.68\% \\
ColBERTv2 & \cellcolor{bestbg}7.38\% & 4.70\% & 19.46\% & 13.09\% & 26.17\% & 17.11\% & 30.87\% & 23.49\% & 13.25\% & 8.99\% \\
BGE-M3 (Hybrid) & 12.75\% & 9.40\% & 32.21\% & 25.50\% & 41.28\% & 35.23\% & 48.32\% & 41.95\% & 21.72\% & 17.55\% \\
Llama-3.1-8B & \cellcolor{worstbg}17.45\% & \cellcolor{worstbg}12.08\% & \cellcolor{worstbg}40.94\% & 28.86\% & \cellcolor{worstbg}47.99\% & 34.90\% & 55.03\% & 38.59\% & \cellcolor{worstbg}27.60\% & 19.43\% \\
Qwen3-Reranker-8B & 16.78\% & 9.40\% & 39.93\% & 27.85\% & 46.64\% & 34.90\% & \cellcolor{worstbg}55.37\% & 39.26\% & 27.17\% & 17.29\% \\
\midrule
& \multicolumn{10}{c}{\textbf{Musical Instruments}} \\
\cmidrule(r){2-11}
\textbf{Model} & \multicolumn{2}{c}{\textbf{Hit@1}} & \multicolumn{2}{c}{\textbf{Hit@5}} & \multicolumn{2}{c}{\textbf{Hit@10}} & \multicolumn{2}{c}{\textbf{Hit@20}} & \multicolumn{2}{c}{\textbf{MRR}} \\
\cmidrule(r){2-3}\cmidrule(r){4-5}\cmidrule(r){6-7}\cmidrule(r){8-9}\cmidrule(r){10-11}
& \textbf{Clean} & \textbf{Error} & \textbf{Clean} & \textbf{Error} & \textbf{Clean} & \textbf{Error} & \textbf{Clean} & \textbf{Error} & \textbf{Clean} & \textbf{Error} \\
\midrule
BM25 & 12.74\% & 8.73\% & 27.57\% & 20.42\% & 34.55\% & 25.83\% & 42.23\% & 31.24\% & 20.43\% & 14.79\% \\
SPLADE & 15.18\% & 11.87\% & 29.32\% & 23.73\% & 35.25\% & 29.32\% & 40.66\% & 34.90\% & 22.39\% & 17.51\% \\
GTE-base & 10.30\% & 5.58\% & 20.07\% & 13.96\% & 25.13\% & 17.45\% & 30.37\% & 21.12\% & 15.51\% & 9.59\% \\
ColBERTv2 & \cellcolor{bestbg}8.20\% & \cellcolor{bestbg}5.24\% & \cellcolor{bestbg}15.88\% & \cellcolor{bestbg}11.17\% & \cellcolor{bestbg}19.55\% & \cellcolor{bestbg}14.66\% & \cellcolor{bestbg}25.83\% & \cellcolor{bestbg}17.10\% & \cellcolor{bestbg}12.14\% & \cellcolor{bestbg}8.47\% \\
BGE-M3 (Hybrid) & 16.58\% & \cellcolor{worstbg}14.31\% & 28.62\% & 23.91\% & 34.03\% & 29.67\% & 41.01\% & \cellcolor{worstbg}36.65\% & 22.70\% & \cellcolor{worstbg}19.67\% \\
Llama-3.1-8B & 17.10\% & 13.61\% & 30.19\% & 23.39\% & 37.70\% & 27.75\% & \cellcolor{worstbg}44.15\% & 31.94\% & 23.62\% & 18.06\% \\
Qwen3-Reranker-8B & \cellcolor{worstbg}17.80\% & 14.14\% & \cellcolor{worstbg}32.98\% & \cellcolor{worstbg}25.65\% & \cellcolor{worstbg}39.62\% & \cellcolor{worstbg}30.72\% & 43.80\% & 32.98\% & \cellcolor{worstbg}24.73\% & 19.18\% \\
\midrule
& \multicolumn{10}{c}{\textbf{Video Games}} \\
\cmidrule(r){2-11}
\textbf{Model} & \multicolumn{2}{c}{\textbf{Hit@1}} & \multicolumn{2}{c}{\textbf{Hit@5}} & \multicolumn{2}{c}{\textbf{Hit@10}} & \multicolumn{2}{c}{\textbf{Hit@20}} & \multicolumn{2}{c}{\textbf{MRR}} \\
\cmidrule(r){2-3}\cmidrule(r){4-5}\cmidrule(r){6-7}\cmidrule(r){8-9}\cmidrule(r){10-11}
& \textbf{Clean} & \textbf{Error} & \textbf{Clean} & \textbf{Error} & \textbf{Clean} & \textbf{Error} & \textbf{Clean} & \textbf{Error} & \textbf{Clean} & \textbf{Error} \\
\midrule
BM25 & 8.89\% & 6.73\% & 18.88\% & 14.06\% & 24.01\% & 17.98\% & 29.53\% & 22.00\% & 14.00\% & 10.45\% \\
SPLADE & 9.74\% & 7.28\% & 20.54\% & 15.72\% & 24.86\% & 20.04\% & 30.14\% & 24.46\% & 15.07\% & 11.57\% \\
GTE-base & \cellcolor{bestbg}5.22\% & 3.57\% & \cellcolor{bestbg}11.50\% & \cellcolor{bestbg}6.88\% & 14.97\% & \cellcolor{bestbg}9.49\% & 18.43\% & \cellcolor{bestbg}12.41\% & \cellcolor{bestbg}8.45\% & 5.58\% \\
ColBERTv2 & 5.83\% & \cellcolor{bestbg}3.06\% & 11.90\% & 7.38\% & \cellcolor{bestbg}14.77\% & 9.79\% & \cellcolor{bestbg}17.68\% & 12.46\% & 8.92\% & \cellcolor{bestbg}5.47\% \\
BGE-M3 (Hybrid) & 10.30\% & \cellcolor{worstbg}8.39\% & 20.49\% & \cellcolor{worstbg}17.28\% & 24.36\% & \cellcolor{worstbg}20.99\% & 29.08\% & \cellcolor{worstbg}25.06\% & 15.26\% & \cellcolor{worstbg}12.73\% \\
Llama-3.1-8B & 9.99\% & 6.83\% & 20.69\% & 15.77\% & 25.92\% & 19.94\% & 30.39\% & 23.20\% & 14.94\% & 10.75\% \\
Qwen3-Reranker-8B & \cellcolor{worstbg}10.45\% & 6.98\% & \cellcolor{worstbg}22.05\% & 16.62\% & \cellcolor{worstbg}26.82\% & 20.29\% & \cellcolor{worstbg}30.44\% & 23.20\% & \cellcolor{worstbg}15.60\% & 11.12\% \\
\bottomrule
\end{tabular}%
}
\usebox{\pqbTableWidthBox}
\par\vspace{2pt}
\normalsize
\begin{minipage}{\wd\pqbTableWidthBox}
\pqbTableNote{In each metric column, the best (highest) value is highlighted with a red background and the worst (lowest) value with a blue background, computed across all seven evaluated models. Clean columns report the personalized syntactic expression query side, and Error columns report the personalized writing-error query side.}
\end{minipage}
\end{table}

\clearpage
\end{document}